%

\input harvmac.tex
\overfullrule=0mm
\hfuzz 15pt

\input epsf.tex
\def\Omit#1{{}}

\font\eightrm=cmr8\font\eighti=cmmi8
\font\eightsy=cmsy8\font\eightit=cmti8
\font\eightsl=cmsl8\font\eighttt=cmtt8\font\eightbf=cmbx8
\font\sixrm=cmr6\font\sixi=cmmi6
\font\sixsy=cmsy6

\font\sixbf=cmbx6

\def\tenpoint{%
\textfont0=\tenrm \scriptfont0=\sevenrm
\scriptscriptfont0=\fiverm \def\rm{\fam0\tenrm}%
\textfont1=\teni \scriptfont1=\seveni
\scriptscriptfont1=\fivei \def\oldstyle{\fam1\teni}%
\textfont2=\tensy \scriptfont2=\sevensy
\scriptscriptfont2=\fivesy
\textfont\itfam=\tenit \def\it{\fam\itfam\tenit}%
\textfont\slfam=\tensl \def\sl{\fam\slfam\tensl}%
\textfont\ttfam=\tentt \def\tt{\fam\ttfam\tentt}%
\textfont\bffam=\tenbf \scriptfont\bffam=\sevenbf
\scriptscriptfont\bffam=\fivebf \def\bf{\fam\bffam\tenbf}%
\abovedisplayskip=12pt plus 3pt minus 9pt
\belowdisplayskip=\abovedisplayskip
\abovedisplayshortskip=0pt plus 3pt 
\belowdisplayshortskip=7pt plus 3pt minus 4pt
\smallskipamount=3pt plus 1pt minus 1pt
\medskipamount=6pt plus 2pt minus 2pt
\bigskipamount=12pt plus 4pt minus 4pt
\setbox\strutbox=\hbox{\vrule height8.5pt depth3.5pt width 0pt} %
\normalbaselineskip=12pt \normalbaselines
\rm}

\def\eightpoint{%
\textfont0=\eightrm \scriptfont0=\sixrm
\scriptscriptfont0=\fiverm \def\rm{\fam0\eightrm}%
\textfont1=\eighti \scriptfont1=\sixi
\scriptscriptfont1=\fivei \def\oldstyle{\fam1\eighti}%
\textfont2=\eightsy 
\scriptscriptfont2=\fivesy
\textfont\itfam=\eightit \def\it{\fam\itfam\eightit}%
\textfont\slfam=\eightsl \def\sl{\fam\slfam\eightsl}%
\textfont\ttfam=\eighttt \def\tt{\fam\ttfam\eighttt}%
\textfont\bffam=\eightbf \scriptfont\bffam=\sixbf
\scriptscriptfont\bffam=\fivebf \def\bf{\fam\bffam\eightbf}%
\abovedisplayskip=9pt plus 2pt minus 6pt
\belowdisplayskip=\abovedisplayskip
\abovedisplayshortskip=0pt plus 2pt 
\belowdisplayshortskip=5pt plus 2pt minus 3pt
\smallskipamount=2pt plus 1pt minus 1pt
\medskipamount=4pt plus 2pt minus 2pt
\bigskipamount=9pt plus 4pt minus 4pt
\setbox\strutbox=\hbox{\vrule height7pt depth2pt width 0pt} %
\normalbaselineskip=9pt \normalbaselines
\rm}

\def\petit{\vskip3mm\eightpoint \skewchar\eighti='177 \skewchar\sixi='177 
\skewchar\eightsy='60 \skewchar\sixsy='60}


\newcount\figno
\figno=0
\def\fig#1#2#3{
\par\begingroup\parindent=0pt\leftskip=1cm\rightskip=1cm\parindent=0pt
\baselineskip=11pt
\global\advance\figno by 1
\midinsert
\epsfxsize=#3
\centerline{\epsfbox{#2}}
\vskip 12pt
{\bf Fig. \the\figno:} #1\par
\endinsert\endgroup\par
}
\def\figlabel#1{\xdef#1{\the\figno}}
\def\encadremath#1{\vbox{\hrule\hbox{\vrule\kern8pt\vbox{\kern8pt
\hbox{$\displaystyle #1$}\kern8pt}
\kern8pt\vrule}\hrule}}

\newwrite\tfile\global\newcount\tabno \global\tabno=1
\def\tab#1#2#3{
\xdef#1{\the\tabno}
\writedef{#1\leftbracket \the\tabno}
\nobreak
\par\begingroup\parindent=0pt\leftskip=1cm\rightskip=1cm\parindent=0pt
\baselineskip=11pt
\midinsert
\centerline{#3}
\vskip 12pt
{\bf Tab. \the\tabno:} #2\par
\endinsert\endgroup\par
\goodbreak
\global\advance\tabno by1
}

%
%

\def\IZ{Z\!\!\!Z}
\def\dC{C\kern-6.5pt I}

\def\Exp{{\rm Exp}}

\def\Ex{{\bf Exercise~}}

 \def\bh{\bar h} \def\bf{\bar f}
\def\bj{\bar j}\def\jb{\bar j}
 \def\bz{\bar z}

\def\IZ{Z\!\!\!Z}

\def\blank#1{}

%
%
\def\frac#1#2{{\scriptstyle{#1 \over #2}}}

\def\ket#1{ | #1 \rangle}
\def\bra#1{ \langle #1 |}
\def\llangle{\langle\!\langle}
\def\rrangle{\rangle\!\rangle}
%
%
\def\CA{{\cal A}}              \def\CC{{\cal C}}
\def\CD{{\cal D}}       \def\CE{{\cal E}}       
       \def\CH{{\cal H}}       \def\CI{{\cal I}}
              
       \def\CN{{\cal N}}       \def\CO{{\cal O}}
\def\CP{{\cal P}}              
\def\CS{{\cal S}}       \def\CT{{\cal T}}       
\def\CV{{\cal V}}              
       
\def\calE{\CE}\def\calH{\CH}
\def\calI{{\cal I}}\def\calN{{\cal N}}\def\calV{{\cal V}}
\def\({ \left( }\def\[{ \left[ }
\def\){ \right) }\def\]{ \right] }
%


\def\IR{\relax{\rm I\kern-.18em R}}
\font\cmss=cmss10 \font\cmsss=cmss10 at 7pt
\def\IZ{\relax\ifmmode\mathchoice
{\hbox{\cmss Z\kern-.4em Z}}{\hbox{\cmss Z\kern-.4em Z}}
{\lower.9pt\hbox{\cmsss Z\kern-.4em Z}}
{\lower1.2pt\hbox{\cmsss Z\kern-.4em Z}}\else{\cmss Z\kern-.4em Z}\fi}
\def\inbar{\,\vrule height1.5ex width.4pt depth0pt}
\def\IB{\relax{\rm I\kern-.18em B}}
\def\ID{\relax{\rm I\kern-.18em D}}
\def\IE{\relax{\rm I\kern-.18em E}}
\def\IF{\relax{\rm I\kern-.18em F}}
\def\IG{\relax\hbox{$\inbar\kern-.3em{\rm G}$}}
\def\IH{\relax{\rm I\kern-.18em H}}
\def\II{\relax{\rm I\kern-.18em I}}
\def\IK{\relax{\rm I\kern-.18em K}}
\def\IL{\relax{\rm I\kern-.18em L}}
\def\IM{\relax{\rm I\kern-.18em M}}
\def\IN{\relax{\rm I\kern-.18em N}}
\def\IO{\relax\hbox{$\inbar\kern-.3em{\rm O}$}}
\def\IP{\relax{\rm I\kern-.18em P}}
\def\IQ{\relax\hbox{$\inbar\kern-.3em{\rm Q}$}}
\def\IGa{\relax\hbox{${\rm I}\kern-.18em\Gamma$}}
\def\IPi{\relax\hbox{${\rm I}\kern-.18em\Pi$}}
\def\ITh{\relax\hbox{$\inbar\kern-.3em\Theta$}}
\def\IOm{\relax\hbox{$\inbar\kern-3.00pt\Omega$}}


\def\d{{\rm d}}

\def\oh{{1\over 2}}
\def\bz{\bar z}

\def\Ga{\alpha}
\def\Gd{\delta}\def\GD{\Delta}\def\GG{\Gamma}

\def\Gl{\lambda}
\def\Gm{\mu}


\def\dim{{\rm dim\,}}

\def\bra{\langle}\def\ket{\rangle}
\def\nind{\noindent}

\def\hepth#1{{\tt hep-th #1}}

\def\slh{\widehat{sl}}

\def\tvp{\vrule height 2pt depth 1pt} 
\def\thp{\vrule height 0.4pt width 0.35em}
\def\cc#1{\hfill#1\hfill}
\setbox34=\hbox{$\scriptstyle {p}$} %
\setbox3=\hbox{$\vcenter{\offinterlineskip
\+ \thp&\cr  
\+ \tvp\cc{}&\tvp\cr 
\+ \thp&\cr  
\+ $\!{}^{\vdots}$\cc{}&$\!{}^{\vdots}$\cr 
\+ \thp&\cr  
\+ \tvp\cc{}&\tvp\cr 
\+ \thp&\cr  }$}
\setbox22=\hbox{$\left.\vbox to \ht3{}\right\}$} 
\def\npbox{n_{\copy3\copy22\copy34}}

\def\\#1 {{\tt\char'134#1} }

\def\a{a}\def\b{b}

\catcode`\@=11
\def\Eqalign#1{\null\,\vcenter{\openup\jot\m@th\ialign{
\strut\hfil$\displaystyle{##}$&$\displaystyle{{}##}$\hfil
&&\qquad\strut\hfil$\displaystyle{##}$&$\displaystyle{{}##}$
\hfil\crcr#1\crcr}}\,}   \catcode`\@=12

\def\encadre#1{\vbox{\hrule\hbox{\vrule\kern8pt\vbox{\kern8pt#1\kern8pt}
\kern8pt\vrule}\hrule}}
\def\encadremath#1{\vbox{\hrule\hbox{\vrule\kern8pt\vbox{\kern8pt
\hbox{$\displaystyle #1$}\kern8pt}
\kern8pt\vrule}\hrule}}

\newdimen\xraise\newcount\nraise
\def\xpoint{\hbox{\vrule height .45pt wNth .45pt}}
\def\udiag#1{\vcenter{\hbox{\hskip.05pt\nraise=0\xraise=0pt
\loop\ifnum\nraise<#1\hskip-.05pt\raise\xraise\xpoint
\advance\nraise by 1\advance\xraise by .4pt\repeat}}}
\def\ddiag#1{\vcenter{\hbox{\hskip.05pt\nraise=0\xraise=0pt
\loop\ifnum\nraise<#1\hskip-.05pt\raise\xraise\xpoint
\advance\nraise by 1\advance\xraise by -.4pt\repeat}}}
\def\bdiamond#1#2#3#4{\raise1pt\hbox{$\scriptstyle#2$}
\,\vcenter{\vbox{\baselineskip12pt
\lineskip1pt\lineskiplimit0pt\hbox{\hskip10pt$\scriptstyle#3$}
\hbox{$\udiag{30}\ddiag{30}$}\vskip-1pt\hbox{$\ddiag{30}\udiag{30}$}
\hbox{\hskip10pt$\scriptstyle#1$}}}\,\raise1pt\hbox{$\scriptstyle#4$}}



\def\IC{\relax\hbox{$\inbar\kern-.3em{\rm C}$}}

\Omit{\def\msy{y }
\message{Do you have the AMS fonts (y/n) ?}\read-1 to \msan
\ifx\msan\msy
\input amssym.def
\input amssym.tex
\def\IZ{\Bbb Z}\def\IR{\Bbb R}\def\IC{\Bbb C}\def\IN{\Bbb N}
\def\II{\Bbb I}\def\IP{\Bbb P}
\def\gg{\goth g}\def\gV{\goth V}
\else \def\gg{g}\def\gV{V} \fi
\def\ggh{\hat\gg}}

\input amssym.def
\input amssym.tex
\def\IZ{\Bbb Z}\def\IR{\Bbb R}\def\IC{\Bbb C}\def\IN{\Bbb N}
\def\II{\Bbb I}\def\IP{\Bbb P}
\def\gg{\goth g}\def\gV{\goth V}
\def\ggh{\hat\gg}

\def\hA{{\hat A}}\def\hD{{\hat D}}\def\hE{{\hat E}}


\lref\BPZ{A.A. Belavin, A.M. Polyakov and A.B. Zamolodchikov, {\it Nucl. 
Phys.} {\bf B 241} (1984) 333-380.}

\lref\KFF{V.G. Kac, {\it Lect. Notes in Phys.} {\bf 94}
(1979) 441-445\semi  B.L. Feigin and D.B.  Fuchs, 
{\it Funct. Anal. and Appl.} {\bf 16} (1982) 114-126; {\it ibid.} 
{\bf 17} (1983) 241-242.}

\lref\Ka{V. Kac, {\it Infinite dimensional algebras}, Cambridge 
University Press; 
V.G. Kac and D.H. Peterson, {\it Adv. Math.} {\bf 53} (1984) 125-264\semi
 J. Fuchs, {\it Affine Lie Algebras and Quantum Groups{}},
Cambridge University Press.  }

\lref\GKO{P. Goddard, A. Kent and D. Olive, {\it Comm. Math. Phys.}
{\bf 103} (1986) 105-119.}

\lref\DFMS{P. Di Francesco, P. Mathieu and D. S\'en\'echal, {\it Conformal 
Field Theory}, Springer Verlag 1997.}

\lref\EV{E. Verlinde, {\it  Nucl. Phys.} {\bf B300} [FS22] (1988) 360-376. }

\lref\Camod{ 
J. Cardy, {\it Nucl Phys} {\bf B 270} (1986) 186-204.}

\lref\Sono{H. Sonoda, {\it Nucl. Phys.} {\bf B281} (1987) 546-572;
{\bf B 284} (1987) 157-192.}

\lref\MS{G. Moore and N. Seiberg,  {\it Nucl. Phys.} {\bf B313}
(1989) 16-40; {\it Comm. Math. Phys.} {\bf 123} (1989) 177-254.}

\lref\DV{E. Verlinde and R. Dijkgraaf, {\it Nucl. Phys. (Proc. Suppl.)}
{\bf 5B} (1988) 87-97.}

\lref\CIZK{A. Cappelli, C. Itzykson and J.-B. Zuber,
{\it Nucl. Phys.} {\bf B280} [FS18] (1987)
445-465; {\it Comm. Math. Phys.} {\bf 113} (1987) 1-26
\semi A. Kato, {\it Mod. Phys. Lett.} {\bf A2} (1987) 585-600.}

\lref\Ga{T. Gannon, {\it Comm. Math. Phys.} {\bf 161} (1994) 233-263;
{\it The Classification of $SU(3)$ Modular Invariants Revisited},
hep-th 9404185.}

\lref\IDG{C. Itzykson, Nucl. Phys. (Proc. Suppl.) {\bf 5B} (1988) 150-165
\semi P. Degiovanni, Comm. Math. Phys. {\bf 127} (1990) 71-99. }

\lref\Gannew{ 
T. Gannon, {\it The monstruous moonshine and the classification of CFT}, 
\hepth{9906167}. }

\lref\BPPZ{R.E. Behrend, P.A. Pearce, V.B. Petkova and J.-B. Zuber, {\sl 
Boundary conditions in RCFT}, {\it Nucl. Phys.} to 
{\bf re}appear, hep-th 9908036.}

\lref\Hazew{M. Hazewinkel, W. Hesselink, D. Siersma and F.D. Veldkamp,
{\it Nieuw Archief voor Wiskunde}, {\bf 3} XXV (1977) 257-307.  }

\lref\Bour{N. Bourbaki, {\it Groupes et Alg\`ebres de Lie}, 
chap. 4--6, Masson 1981; 
J.E. Humphreys, {\it Introduction to Lie Algebras and Representation
Theory}, Springer Verlag, New York, 1972.}

\lref\Coxhum{
H.S.M. Coxeter, {\it Ann. Math.} {\bf 35} (1934) 588-621\semi
J.E. Humphreys, {\it Reflection Groups and Coxeter Groups}, 
Cambridge Univ. Pr. 1990. }

\lref\Slo{For a review and references, see P. Slodowy, {\it Platonic solids, 
Kleinian singularities and Lie groups}, Lect. Notes in Math. {\bf 1008}
(1983) 102-138, Springer; {\it Algebraic groups 
and resolutions of Kleinian singularities},
RIMS-1086, Kyoto 1996.}

\lref\Arn{V.I. Arnold, S.M. Gusein-Zaide and A.N. Varchenko, 
{\it Singularities of differential maps}, Birkh\"auser, Basel 1985.}

\lref\Gab{P. Gabriel, {\it Manusc. Math.} {\bf 6} (1972) 71-103.}

\lref\GHJ{F.M. Goodman, P. de la Harpe and V.F.R. Jones, {\it Coxeter
Graphs and Towers of Algebras}, Springer-Verlag, Berlin (1989).}

\lref\Hille{E. Hille, {\it Ordinary Differential Equations in 
the Complex Domain}, {J. Wiley, 1976}.}

\lref\Jo{V.F.R. Jones, {\it Invent. Math.} {\bf 72} (1983) 1-25.}

\lref\McK{J. McKay, {\it Proc. Symp. Pure Math. }{\bf 37} (1980) 183-186.}

\lref\Bry{J.-L. Brylinski, {\sl A correspondance dual to McKay's}, 
{\tt alg-geom 9612003}.}

\lref\WDVV{
T. Eguchi and S.K. Yang, {\it Mod. Phys. Lett.} {\bf A5} (1990) 1693-1701
\semi
R. Dijkgraaf, E. Verlinde and H. Verlinde, Nucl. Phys. 
{\bf B352} 59-86 (1991):
in {\it String Theory and Quantum Gravity},
proceedings of the 11990 Trieste Spring School, M. Green et al. {\it eds.},
World Sc. 1991.}

\lref\VPun{V. Pasquier, {\it Nucl. Phys.} {\bf B285} [FS19] (1987) 162-172
\semi V. Pasquier, {\it J. Phys.} {\bf A20} (1987) 5707-5717.}

\lref\LVWM{
 W. Lerche, C. Vafa and N. Warner,
{\it Nucl. Phys. } {\bf B 324} (1989) 427-474
\semi E. Martinec, in 
{\it Criticality, 
catastrophes and compactifications}, in 
{\it Physics and mathematics of strings}, V.G. Knizhnik memorial volume,
L. Brink, D. Friedan and A.M. Polyakov eds., World Scientific 1990
\semi
P. Howe and P. West, {\it Phys. Lett.} {\bf B223} (1989) 377-385; 
{\it ibid.} {\bf B227}  (1989) 397-405.
}

\lref\Zubkyoto{J.-B. Zuber, {\sl C-algebras and their applications to
reflection groups and conformal field theories}, lectures at RIMS, 
Kyoto December 1996, {\tt hep-th 9707034}. }

\lref\OcnEK{A. Ocneanu, ``Quantized groups, string algebras and Galois theory
for algebras'', in {\it Operator Algebras and Applications},  D. Evans and
M. Takesaki eds, 1988, pp. 119-172
\semi D. Evans and Y. Kawahigashi,
{\it Publ. RIMS, Kyoto Univers. } {\bf 30} (1994) 151-166.}

\lref\Pasq{V. Pasquier, {\sl Mod\`eles Exacts Invariants Conformes}, 
Th\`ese d'Etat, Orsay, 1988.}

\lref\DFZun{P. Di Francesco and J.-B. Zuber,
{\it Nucl. Phys.} {\bf B338} (1990) 602-646
.}

\lref\Soch{N. Sochen, {\it Nucl. Phys.} {\bf B360} (1991) 613-640. }

\lref\PZdeu{V.B. Petkova and J.-B. Zuber, {\it Nucl. Phys.} {\bf B 463} (1996) 161-193, \hepth{9510175}; \hepth{9701103}. }

\lref\Ocn{A. Ocneanu, {\it Paths on Coxeter Diagrams},
in {\it Lectures on Operator Theory}
Fields Institute Monographies, Rajarama Bhat et al edrs, 
AMS 1999.}

\lref\BEK{ 
{J. B\"ockenhauer  and D.E. Evans, } 
{\it Comm.Math. Phys.} 
{\bf 197} (1998) 361-386; {\it ibidem} {\bf 200} 
(1999) {57-103}, \hepth{9805023}; {\it ibidem} {\bf 205} (1999) 183-228,  
{\tt hep-th/9812110};
J. B\"ockenhauer, D.E. Evans, and Y. Kawahigashi, 
{\it Comm.Math. Phys.} {\bf 208} (1999) 429-487, {\tt math-OA 9904109};
{\it ibidem} {\bf 210} (2000) 733-784;  
see also D. Evans' lectures at this school.}

\lref\Xu{F. Xu, {\it Comm. Math. Phys.} {\bf 192} (1998) 349-403. }

\lref\AO{A. Ocneanu, lectures at this school.}

\lref\Ztani{J.-B. Zuber, {\it Generalized Dynkin diagrams and root systems
and their folding}, Proccedings of the Taniguchi meeting, Kyoto Dec 1996, 
{\it Topological Field Theory, Primitive Forms and Related Topics},
M. Kashiwara, A. Matsuo, K. Saito and I. Satake edrs, Birkh\"auser.}

\lref\GZV{S.M. Gusein-Zaide and A.N. Varchenko, {\it Verlinde algebras 
and the intersection form on vanishing cycles}, \hepth{9610058}.} 

\lref\RS{
A. Recknagel and V. Schomerus, {\it Nucl. Phys.} {\bf B 531}
{(1998)} {185-225}
.}

\lref\FS{J. Fuchs and C. Schweigert, {\it Nucl. Phys.} {\bf B 530 } (1998) 99-136.}

\lref\Cabc{J.L. Cardy, {\it Nucl. Phys.} {\bf 324} (1989) {581-596}.}

\lref\PSS{{G. Pradisi, A. Sagnotti and Ya.S. Stanev}
{\it Phys. Lett.}{\bf 381} {(1996)} {97-104}.}

\lref\PaSa{V. Pasquier and H. Saleur {\it Nucl. Phys. } {\bf B 330} (1990)
523-556.}

\lref\Kos{B. Kostant, {\it Proc. Natl. Acad. Sci. USA} {\bf 81} (1984) 
5275-5277;  
{{\it Ast\'erisque (Soci\'et\'e Math\'ematique de France)},}{ }(1988) {209-255, }
{\it The McKay Correspondence, the Coxeter Element and Representation
Theory}.}

\lref\Dor{
P. Dorey, {\it Int. J. Mod. Phys.}{\bf 8} (1993) 193-208.}

\lref\BCDS{H. Braden, E. Corrigan, P.E. Dorey  and R. Sasaki,  {\it Nucl. Phys.}
{\bf B 338} (1990) {689-746}. }

\lref\MCOBSS{
B.M. McCoy and W. Orrick {\it Phys. Lett. }{\bf A 230}
(1997) {24-32} 
\semi
{M.T. Batchelor and K.A. Seaton, K.A., }{\tt cond-mat 9803206},  
{\sl Excitations in the
diluate $A_L$ lattice model: $E_6$, $E_7$ and $E_8$ mass spectra}
\semi
{J. Suzuki}, {\tt cond-mat 9805241}, {\sl Quantum Jacobi-Trudi Formula and
$E_8$ Structure in the Ising Model in a Field}.}

\lref\BI{E. Bannai, T. Ito, {\it Algebraic Combinatorics I: Association
Schemes}, Benjamin/Cummings (1984).}

\lref\DFZdeu{P. Di Francesco and J.-B. Zuber,
in {\it Recent Developments in Conformal Field Theories}, Trieste Conference
1989, S. Randjbar-Daemi, E. Sezgin and J.-B. Zuber eds., World Scientific
1990 \semi
P. Di Francesco, {\it Int. J. Mod. Phys.} {\bf A7} (1992) 407-500.}

\lref\BP{R.E. Behrend and P.A. Pearce, {\it J. Phys. A} {\bf 29} (1996) 
7827-7835; {\it Int. J. Mod. Phys.} {\bf 11} (1997) 2833-2847; 
{\it Integrable and Conformal Boundary Conditions for sl(2) A-D-E 
Lattice Models and Unitary Minimal Conformal Field Theories},
 {\tt hep-th hep-th/0006094}.}


\def\today{\number\day\ \ifcase\month\or January \or February \or March \or
April \or May \or June \or July \or August \or September \or October \or
November \or December\fi\space\number\year}   


\Title
{SPhT 00/072}
{{\vbox {
\vskip-10mm
\centerline{ CFT, BCFT, ADE }
\medskip
\centerline{and all that}
}}}

\medskip
\centerline{J.-B. Zuber}
\medskip\centerline{Service de Physique Th\'eorique}
\medskip\centerline{CEA Saclay}
\medskip\centerline{F 91191 Gif-sur-Yvette Cedex, France}
\vskip .2in

\noindent 
These pedagogical lectures  present some material, classical 
or more recent, on (Rational) Conformal Field Theories and 
their general setting ``in the bulk'' or in the presence of
a boundary. Two well posed problems are
the classification of modular invariant partition functions and
the determination of  boundary conditions consistent with conformal 
invariance. It is shown why the two problems are intimately connected
and how graphs --ADE Dynkin diagrams and their generalizations--
appear in a natural way.

\bigskip

\Date{\vbox{\leftline{ Lectures at ``Quantum Symmetries in Theoretical
Physics and Mathematics'', }
\leftline{ January 10--21, 2000, Bariloche, Argentina.}}}
%

\secno=-1
\newsec{Introduction}
\nind
These lectures aim at presenting some curious features encountered
in the study of 2~D conformal field theories. The key words are
graphs, or Dynkin diagrams, and indeed we shall encounter new
avatars of the ADE Dynkin diagrams, and some generalizations thereof.
The first lecture is devoted to a lightning review of Conformal
Field Theory (CFT), essentially to recall essential notions and to establish
basic notations. The study of modular invariant partition functions
for theories related to the simplest Lie algebra $sl(2)$ leads
to an ADE classification, as has been known for more than ten years.
A certain frustration comes from the fact that we have no
good reason to explain why this ADE classification appears,
or no definite way to connect it to another existing classification
(Lecture 2). Or, which amounts to the
same, we have too many: depending on the way we look at these
$sl(2)$ theories --their topological counterparts, their
lattice realization--
the reason looks different.
Moreover, when we turn to higher rank algebras, 
the situation is even more elusive: it had been guessed long ago
that classification should involve again graphs, though for reasons
not very well understood,  and a list of graphs 
had been proposed in the case of $sl(3)$.
Recent progress has confirmed these expectations: 
as will be discussed in the final lecture 3,
through the study of boundary conditions  
we now understand why graphs are naturally associated with
CFTs  and which properties they must satisfy. In the case
of $sl(2)$, this leads in a straightforward way (but up to
a little subtlety)  to the ADE diagrams.
Moreover, the classification of the graphs relevant in the case of $sl(3)$ 
has just been completed, see A. Ocneanu's lectures at this school.


\newsec{A crash course on CFT}
\nind
This section is devoted to a fast summary of  concepts and notations
in conformal field theories (CFTs).

\subsec{On CFTs}
\nind
A conformal field theory  is a quantum field theory endowed with
covariance properties under conformal transformations. We shall
restrict our attention to two
dimensions, and in a first step to the Euclidean plane,
 where the conformal transformations are realized by any analytic change of
coordinates $z\mapsto \zeta(z),\ \bar z\mapsto \bar\zeta(\bar z)$. In fact,
it appears that the contributions of the variables $z$ and $\bar z$
decouple~{\BPZ} and that $z$ and $\bar z$ may be regarded as
independent variables.
The $zz$ component of the energy-momentum tensor $T_{\mu\nu}$
is analytic as a consequence of its tracelessness and conservation,
and denoted $T(z)$,  $T(z):= T_{zz}(z)$,
and likewise $\bar T(\bar z):=T_{\bar z\bar z}(\bar z)$ is antianalytic.
$T(z)$ is the  generator of the infinitesimal change
$z\mapsto \zeta=z +\epsilon(z)$
and likewise $\bar T(\bar z)$ for $\bar z\mapsto \bar\zeta(\bar z)$,
in the sense that under such a change a correlation function of
fields undergoes the change 
\eqn\wardid
{\delta\bra\phi_{i_1}(z_1,\bar z_1)\cdots\phi_{i_n}(z_n,\bar z_n)\ket
={1\over 2\pi i}\oint_\CC \, \d w\, \epsilon(w)\,
\bra T(w) \phi_{i_1}(z_1,\bar z_1)\cdots\phi_{i_n}(z_n,\bar z_n)\ket
+ {\rm c.c.}
}
with an integration contour encircling all points $z_1,\cdots, z_n$.

Fields of a CFT are assigned an explicit  transformation under these changes
of variables.
In particular, {\it primary fields} transform as $(h,\bh)$ forms,
i.e. according to
\eqn\prim
{\tilde\phi(\zeta,\bar\zeta)=\Big({\partial z\over \partial \zeta}\Big)^h
\Big({\partial \bz\over \partial \bar \zeta}\Big)^{\bar h} \phi(z,\bar z)}
where the real numbers $(h,\bar h)$ are the {\it conformal weights} of
the field $\phi$ (see below for a representation-theoretic interpretation).
For an infinitesimal change, $\zeta=z+\epsilon(z)$, 
$\bar\zeta=\bar z+\bar\epsilon(\bar z)$,  
if we set $\tilde\phi(z,\bz)=\phi(z,\bz)-\delta\phi(z,\bz)$
\eqn\infinit
{ \delta \phi= (h \epsilon' + \epsilon \partial_z )\phi + {\rm c.c.}}
{\petit  \noindent
Here and in the following, ``c.c.'' denotes the formal complex conjugate,
$(z,h)\to (\bz,\bar h)$, 
``formal'' because $\bar z$ is at this stage independent of $z$ and
$h$ and $\bar h$ are also a priori independent. When the condition that
$\bar z=z^*$ is imposed, 
$h+\bar h$ turns out to be the scaling dimension of the field $\phi$
and $h - \bar h$ its spin: locality (singlevaluedness of the correlators)
imposes only on $h-\bar h$ to be an integer or half-integer.}

In contrast to primary fields, the change of $T(z)$ itself has the form
\eqn\varT
{ {\tilde T}(\zeta)=\Big({\partial z\over \partial\zeta} \Big)^2
 T(z) +{c\over 12} \{z,\zeta\} }
where $\{z,\zeta\} $ denotes the schwarzian derivative
\eqn\schwarz
{ \{z,\zeta\}=
{{\partial^3 z\over\partial \zeta^3}\over
{\partial z\over\partial \zeta}} 
-{3\over 2} \Bigg({{\partial^2 z\over\partial \zeta^2}\over
{\partial z\over\partial \zeta}} \Bigg)^2 }
and where  the parameter $c$ in front of the anomalous schwarzian term
is the {\it central charge} (of the Virasoro
algebra, to come soon!). In other words, $T$ transforms almost as a
primary field of conformal weights $(h,\bar h)=(2,0)$ --a conserved
current of scaling dimension 2-- up to the schwarzian anomaly.
\par\nind {\petit \Ex: derive the form of the variation of $T$ under
an infinitesimal transformation, i.e. the analog of \infinit. }

\bigskip
In the spirit of local field theory, we assume that correlation functions
as above are well defined in the complex plane with possible
singularities only at coinciding points $z_i=z_j$ or $w=z_i$. Close to
these points, there is a short distance expansion of products of
fields.
As an exercise, using Cauchy theorem, show that equation \infinit\
(and its analog for $T$)
may be rephrased as a statement on the expansions
\eqna\ope
$$ \eqalignno{ T(w) \phi(z,\bz)&= {h \phi(z,\bz)\over (w-z)^2}
+ {\partial \phi(z,\bz)\over (w-z)}+{\rm regular} & \ope a
\cr
T(w) T(z) &= {{c\over 2}\over (w-z)^4}+ { 2 T(z)\over (w-z)^2}
+{\partial T(z)\over (w-z)}+{\rm regular} & \ope b \cr}$$
with similar expressions for the products with $\bar T(\bz)$.
Eqs \ope{} are meant to describe the singular behaviour of
correlation functions $\bra T(w) \phi(z,\bz)\cdots\ket$, 
$\bra T(w) T(z)\cdots\ket$, in the presence of spectator fields, 
as $z\to w$. 

\bigskip
As usual in Quantum Field Theory, 
it is good to have two dual pictures at hand: the one
dealing with correlation (Green) functions, as we have done so far,
  and in the spirit of
quantum mechanics, the operator formalism, which
describes the system by ``states'', i.e.
vectors in a Hilbert space. 
In CFT, it is appropriate to
think of a radial quantization in the plane: surfaces of equal ``time''
are circles centered at the origin, and the Hamiltonian is the dilatation
operator. The origin in the plane plays the r\^ole of
remote past, the remote future lies on the circle at infinity.
On any circle, there is a description of the system in terms of a
Hilbert space $\cal H$ of states, on which field {\it operators} act.
If we expand the energy momentum tensor on its Laurent modes
\eqn\laurent
{T(z)= \sum_{n=-\infty}^\infty  z^{-n-2} L_n }
it is a good exercise (making use again of Cauchy theorem) to check that
the expansion \ope{b}, now regarded as an ``operator product expansion'' (OPE),
may be rephrased as a commutation relation between the $L$'s
\eqn\virasoro
{[L_n,L_m] = (n-m) L_{n+m} +{c\over 12} n(n^2-1) \delta_{n+m,0}\ .}
This is the celebrated {\bf Virasoro algebra}, in which the {\it central 
charge} $c$ appears indeed as the coefficient of the central term.
Note that $L_0$ is the generator of dilatations, $L_{-1}$ the one of
translations. Together with $L_1$, they form a subalgebra.
\par\nind{\petit What is the
interpretation of $L_1$ and of that subalgebra?}

\par\nind{\petit \Ex: derive the commutation relation of $L_n$ with the
field operator $\phi$, as a consequence of the equation \ope{a}. }

\subsec{Extension to other chiral algebras.}
\nind
A frequently encountered situation in CFT is that there
is a larger (``extended'') {\it chiral} algebra $\cal A$
encompassing the Virasoro algebra, and acting on fields of the theory.
The latter thus fall again in representations of $\cal A$.
The most common cases are those involving a current algebra, i.e.
the affine extension $\ggh$ of a Lie algebra $\gg$, or the
so-called $W$ algebras, or the various superconformal algebras, etc.
\par\nind {\petit In the affine algebra $\gg$, the important objects from our
standpoint
are the current $J(z)$ or its moments $J_n$, with values in the
adjoint representation of $\gg$. In some basis, they satisfy the
commutation relation
\eqn\comrel{[J_n^a,J_m^b]=i f^{ab}{}_c J_{n+m}^c +{{\hat k}}\, n\, \delta_{n+m,\,  0}
\,\delta_{ab}\ ,}
with $\hat k$ a central element.}
\par \nind The Virasoro algebra is
either a proper subalgebra (for ex, in the superconformal cases), or
contained in the enveloping algebra
of this chiral algebra. For example, in affine algebras,
the energy momentum tensor and Virasoro generators are obtained
through the Sugawara construction as quadratic forms in the currents:
$ T(z)= {\rm const. }\  
 :({\bf J}(z))^2:$. (The colons refer to a specific regularization
of this ill-defined product of two currents at coinciding points, and
the constant is fixed in any irreducible representation\dots).
For many cases of such an extended chiral algebra,
the representation theory has been developed (maybe in
less detail for $W$ algebras, even less for other,  more exotic,
extended algebras \dots).


\subsec{Elements of Representation theory of the Virasoro algebra}
\nind
The representations that we shall consider are the {\it highest weight
representations (or Verma modules)}. They are parametrized by a real number
$h$: the Verma module $\gV_h$
is generated from a highest weight (h.w.) vector denoted $|h\ket$
satisfying
\eqn\hw
{L_0|h\ket =h |h\ket \qquad \qquad L_{n}|h\ket =0,\quad \forall n\in \IN}
by the action of the $L_{-n}$, $n>0$
\eqn\span
{\gV_h={\rm Span}\{L_{-p_1} L_{-p_2}\cdots L_{-p_r} |h\ket \},
\quad 1\le p_1\le p_2\le\cdots\le p_r\, .}
Such a representation has the important property of being graded
for the action  of the Virasoro generator $L_0$. This means
that the spectrum of $L_0$ in $\gV_h$ is of the form $\{h,h+1,h+2, 
\cdots\}$.
The subspace of eigenvalue $h+N$ is called the eigenspace of level $N$.

\medskip
Whether this module is or is not irreducible is the object of a theorem
(Kac, Feigin-Fuchs)\KFF\par\nind
\smallskip
{\bf Theorem}\ {\sl Let $c=1-6/x(x+1)$, where $x\in \IC$. Then $\gV_h$
is reducible iff there exist two positive integers $r$ and $s$
such that
\eqn\reduci
{ h= h_{rs}:={\((r(x+1)-s x\)^2-1\over 4x(x+1)}\ . }}
If $h$ takes one of these values, then $\gV_h$ is reducible in the sense
that it contains a ``singular''
vector, i.e. a vector satisfying the axioms \hw\ of
a h.w. vector. This vector thus supports itself a h.w. module,
which is a submodule of $\gV_h$. Moreover the theorem asserts that
this ``degeneracy'' occurs at level $r.s$.
 In fact $\gV_h$ may contain several
such submodules,  with a non trivial intersection: this is what happens
if the parameter $x$ is rational.
One constructs the irreducible representation $\CV_h$ by quotienting out
this (or these) submodule(s) of $\gV_h$.
\par\nind
\vskip-2mm
{\petit Assume that the parameter $x$ in \reduci\ is of the form
$p'/(p-p')$, with $p,p'$ two coprime integers. 
Show that $h=h_{r\,s}=h_{p'-r\,p-s}$
so that there are degeneracies at the two levels $r.s$ and $(p'-r).
(p-s)$ and thus two distinct submodules in $\gV_h$.}\par\nind
\medskip

Highest weight representations of other chiral algebras
may also be constructed.
For example, let us sketch the results for an affine algebra $\ggh$
associated with a simple algebra $\gg$ \Ka.
Let $\alpha_1,\cdots,\alpha_r$ and $\theta$ be the ordinary simple roots 
and the highest root of the finite
algebra $\gg$,  $\alpha_i^\vee=2 \alpha_i/(\alpha_i,\alpha_i)$
the corresponding coroots; $\theta$ is normalised by $(\theta,\theta)=2$.
Then the so-called {\it integrable} representations of the algebra $\ggh$
are labelled by a pair $(\bar\lambda,k)$ where $k$ is a non negative 
integer (the ``level'' of the representation) and where  
$\bar\lambda\in P$, ($P$ the weight lattice of $\gg$),  
 is subject to the inequalities
\eqn\Walcov{(\bar\lambda, \alpha^\vee_i) \in \IN 
,\quad i=1,\cdots ,r\ ,\qquad 
(\bar\lambda,\theta)\le k\ .}
Many formulae simplify if
expressed in terms of the shifted weight $\lambda=\bar\lambda+\rho$,
with $\rho$ the Weyl vector $\rho=\oh\sum_{\alpha>0} \alpha$. In the case of
$\gg=sl(N)$, $r=N-1$, these formulae reduce to $\theta=\sum_1^{N-1} \alpha_i$,
the shifted weight
$\lambda=\sum_{i=1}^{N-1} \lambda_i \Lambda_i$, with
$\Lambda_i$ the dominant fundamental weights of $sl(N)$,
 satisfies the inequalities
$\lambda_i\ge 1, \sum_{i=1}^{N-1}\lambda_i \le k+N-1$.
%
 The simplest example is of course that of
$\slh(2)$ for which the representations are labelled by the pair of integers
$(\lambda,k)$, with $1 \le \lambda \le k+1$, ($\lambda$ may be thought of
as $2j+1$, i.e. the dimension of the corresponding finite-dimensional
spin $j$ representation of $sl(2)$).

\medskip
If $\CV_i$ is some representation of a chiral algebra, we shall use the 
label  $i^*$ to denote
the complex conjugate representation. It may happen that $\CV_{i^*}$ is 
identical or equivalent to $\CV_i$, like for Vir, or $\slh(2)$.
We shall in general label by $i=1$ the identity representation; its
conformal weight (eigenvalue of $L_0$ on the highest weight vector)
vanishes.

As in the case of Vir discussed above, these representations are graded
for the action of the Virasoro generator $L_0$. 
The spectrum of $L_0$ in $\CV_i$ 
is of the form $\{h_i,h_i+1,h_i+2, \cdots\}$, with non-trivial
multiplicities $\#_n= \dim$(subspace of eigenvalue $h+n$). It is thus
natural to introduce a generating function of these multiplicities,
i.e. a function of a dummy variable $q$, the {\it character} of the
representation $\CV_i$
\eqn\chara
{\chi_{i}(q)=\tr_{\CV_{i}}  q^{L_0-{{c\over 24}}}=q^{h_i-{{c\over 24}}}
\sum_{n=0}^\infty \#_n q^n \ . }
\nind{
Show that in the original h.w. Verma module $\gV_h$ of Vir,
the character is simply $\chi_h(q)={q^{h-{c\over 24}}\over \prod_1^\infty
(1-q^n)}$. Assuming that for  $c=1$,  the  representations with a
singular vector
have  a conformal weight given by the limit $x\to \infty$ of \reduci,
namely $h={\ell^2/ 4}$, $\ell\in \IN$, $r=\ell+1$, $s=1$, show that
$\chi_h(q)={q^{\ell^2\over 4}(1-q^{\ell+1})\over \eta(q)}$ with
Dedekind's eta function:
 $\eta(q)=q^{{1\over 24}} \prod_1^\infty (1-q^n)$.
Another interesting case is for $c<1$, when the parameter $x$ is
rational: one writes
\eqn\minrep
{c=1-{6(p-p')^2\over pp'},\quad p,p'\in \IN}
and one  concentrates on the  irreducible representations (``irrep'')
with $h$ of the form
\eqn\hmin{h_{rs}=h_{p'-r,p-s}={(rp-sp')^2-(p-p')^2\over 4 pp'}, \qquad 1\le r\le p'-1,
\ 1\le s \le p-1\ .}
The character of this irrep reads, with $\lambda:=(rp-sp')$,
$\lambda':=(rp+sp')$
\eqn\chirrep
{\chi_{rs}={1\over \eta(q)}
\sum_{n\in \IZ} \( q^{(2npp'+\lambda)^2\over 4pp'}-
q^{(2npp'+\lambda')^2\over 4pp'}\) \ . }
}

\medskip
In representations of the affine algebra $\ggh$, we may also consider
the characters $\tr\, q^{L_0-c/24}$. They are called ``specialized 
characters''
since they count states according to their $L_0$ grading only.
Non-specialized characters can be introduced, which are sensitive to
generators of the Cartan subalgebra ${\bf J}_0$
\eqn\spech{ \chi(q,{\bf u})= \tr\, q^{L_0-{c\over 24}} e^{2\pi i ({\bf u},{\bf J}_0)}
\ .}
For the case of $\slh(2)$, and for the representations $(\lambda, k)$
discussed  above
\eqn\sldch{\chi_\lambda(q) ={1\over \eta^3(q) }\sum_{p=-\infty}^\infty
(2(k+2)p+\lambda)\,q^{{(2(k+2)p+\lambda)^2\over 4(k+2)}}\ .}
The expressions of non-specialized characters and for general
algebras may be found in \Ka.

\medskip
Similar considerations apply to other chiral algebras\dots

\subsec{Modular properties of characters}
\nind If the multiplicity $\#_n$ doesn't grow too fast in \chara,
this sum converges for $|q|<1$: it
is thus natural to write $q=\exp 2i\pi \tau$, with $\tau$ a complex number
in the upper half-plane.

It is a remarkable fact that such functions $\chi$ enjoy beautiful
transformation properties under the action of the modular group on the
variable $\tau$
\eqn\modgr{ \pmatrix{a&b\cr c&d\cr} \in PSL(2,\IZ)\quad :
\tau \mapsto {a\tau+b\over c \tau +d} }
(i.e. $a,b,c,d$ integers defined up to a global sign, with
$ad-bc=1$).

By definition,  Rational Conformal Field Theories
(RCFT) are CFTs that are consistently described by 
a finite set $\CI$  of representations $\calV_i$, 
$i\in \cal I$, of a certain chiral algebra $\CA$. 
Moreover the corresponding characters $\chi_i(q)$
form a finite dimensional unitary representation of the modular group
 (in fact of its double covering, see below): they
transform among themselves linearly (and unitarily) under the action
of \modgr. It is well known
 that the modular group is generated by the two transformations
\eqn\modtr{
T\ :\ \tau\mapsto \tau+1\qquad\quad S\ :\ \tau\mapsto-{1\over\tau}\ .}
It is clear from the definition \chara\ that under $T$
\eqn\Ttrans
{\chi_i(q)\to \chi_i(q\, e^{2i\pi }) =e^{2i\pi(h_i-{c\over 24})} \chi_i(q)}
and the non-trivial part of the above statement is  that,
if $\tilde q:= \exp -{2i\pi\over \tau}$ there exists
a unitary $|\CI|\times |\CI|$ matrix $S$ such that
\eqn\Strans
{\chi_i( q) =\sum_{j\in \CI} S_{ij} \chi_j(\tilde q)\ . }
Moreover the matrix $S$ satisfies
$S^T=S$,\quad $S^\dagger=S^{-1}$,\quad $(S_{ij})^*=S_{i^*j}
=S_{ij^*}$, $S^2=C={\rm the\ conjugation\ matrix}$
defined by $C_{ij}=\delta_{ij^*}$, \quad $S^4=I$.
\footnote{${}^{(2)}$}{The fact that $S^2=C$ rather than $S^2=I$
as expected from the transformation \modtr\ signals that we
are dealing with a representation of a double 
covering of the modular group.}

\nind
{\bf 1st Example}: for the $c<1$ minimal representations,
$$\calI=\{(r,s)\equiv (p'-r, p-s)\ ;\quad  1\le r\le p'-1,\ 1\le s\le p-1\}$$
with $h_{rs}$ given in \hmin, the $S$ matrix reads
\eqn\Smin
{S_{rs,r's'}=\sqrt{{8\over pp'}}(-1)^{(r+s)(r'+s')}
\sin\pi rr'{p-p'\over p'}\,\sin\pi ss'{p-p'\over p} }
\Omit{(It may be shown that this is the only case of rational CFTs
involving only a finite number of representations of the Virasoro
algebra itself, whence the name ``minimal'' attached to theories
constructed out of these representations.) }

\nind {\bf 2nd Example}: the $\slh(2)$ affine algebra\par\nind
At level $k$, we have seen that the integrable representations
are labelled by the set
$ \calI=\{1,2,\cdots, k+1\}$. The conformal weight
of  representation $(\lambda)$ reads $h_\Gl= (\Gl^2 -1)/4(k+2)$,
$ \Gl\in \calI$.
 Then one finds
\eqn\Ssld{S_{\Gl\Gm} = \sqrt{2\over k+2} \sin{\pi \Gl\Gm\over k+2}\ ,\quad
\Gl,\Gm\in \calI\,\ .}
For non-specialised characters, the transformation reads:
\eqn\trunsp{\chi_\Gl(q,{\bf u})=
e^{i k\pi ({\bf u},{\bf u})/\tau} \,
 \sum_{\Gm\in  \CI} S_{\Gl\Gm} \chi_\Gm(\tilde q, {\bf u}/\tau) \ .}
The expression for more general affine algebras may be found
in \Ka.

\nind{\petit One notes that the $S$ matrix of the minimal case \Smin\
is ``almost'' the tensor product of two matrices of the
form \Ssld, at two different  levels $k=p-2$
and $k'=p'-2$. This would be true for $|p-p'|=1$, and 
if one could omit the
identification $(r,s)\equiv (p'-r, p-s)$. This is of course not a
coincidence but reflects the ``coset construction'' of $c<1$
representations of Vir out of the affine algebra $\slh(2)$ \GKO. 
 }


\subsec{Notion of Fusion Algebra}
\nind
The last concept of crucial importance for our discussion is
that of fusion algebra. Fusion is an operation among
representations of chiral algebras of RCFTs, inherited from the operator
product algebra of Quantum Field Theory. 
It looks similar to the usual tensor product of
representations, but contrary to the latter, it is consistent 
with the finiteness of the set $\CI$ and it preserves the
central elements (instead of adding them). I shall refer to the
literature \refs{\Ka,\DFMS} for a systematic discussion of this concept,
and just introduce a notation $\star$ to denote it and  distinguish
it from the tensor product. It is natural to decompose the
fusion of two representations of a chiral algebra on the irreps,
thus defining multiplicities, or ``fusion coefficients''
\eqn\fuscoef
{  \CV_i \star \CV_j =\oplus_k\, \calN_{ij}{}^k\, \calV_k,
\qquad  \calN_{ij}{}^k \in \IN\ .}

There is a remarkable formula, due to Verlinde \EV, expressing these
multiplicities in terms of the unitary  matrix $S$:
\eqn\verl
{ \calN_{ij}{}^k=\sum_{\ell\in\calI}{S_{i \ell}\, S_{j\ell}
\left(S_{k\ell}\right)\!{}^*
\over S_{1\ell}}}

{\petit 
Note that the fact that the r.h.s. of \verl, computed with the matrices
\Ssld\ or \Smin\ yields non negative integers is a priori non-trivial!}

\medskip

This completes our review of the basics of RCFTs. The data
${c}$  (or $k$, etc), $\calV_i, h_i, i\in\calI$, $S_i{}^j$,
${\CN}_{ij}{}^k$, form what I call the ``chiral data'': they  are relative
to a ``chiral half'' of the conformal theory (in the plane), which means
they refer only to the holomorphic variables $z$
or to their antiholomorphic counterparts $\bz$,
rather than to the pair $(z,\bz)$.

Our task is now to use these ingredients to construct physically
sensible theories.


\newsec{Modular Invariant Partition Functions}

\nind
In the plane punctured at the origin, equipped with the coordinate $z$,
or equivalently on the cylinder of perimeter $L$ with the coordinate $w$,
with the conformal mapping from the latter to the former $z=\exp -2\pi i w/L$,
a given RCFT is described by a Hilbert space $\CH_P$. This Hilbert space is
decomposable into a {\it finite} sum of irreps of {\bf two} copies
of the chiral algebra (Vir or else), associated with the holomorphic and
anti-holomorphic ``sectors'' of the theory:
\eqn\hilbert{
\CH_P=\oplus N_{j\bar j} \calV_j\otimes \calV_{\bar j}\ , }
with (non negative integer) multiplicities $N_{j\bar j}$.
On the cylinder, it is natural to think of the Hamiltonian as the
operator of translation along its axis (the imaginary axis in $w$),
or along any helix, defined by its period   $\tau L$ in the $w$ plane,
with $\Im m\,\tau>0$.
If $L_{-1}^{{\rm cyl}}$
and  $\bar L_{-1}^{{\rm cyl}}$ are the two generators of translation
in $w$ and $\bar w$, $H^{{\rm cyl},\tau}= (\tau L_{-1}^{{\rm cyl}}
+\bar \tau \bar L_{-1}^{{\rm cyl}})$.
Mapped back in the plane using the transformation law of the
energy-momentum tensor \varT, $ L_{-1}^{{\rm cyl}}$ reads
\eqn\cyltopl{ L_{-1}^{{\rm cyl}} = -{2\pi i\over L} ( L_0 -{c\over 24})}
where the term $c/24$ comes from the schwarzian derivative of the
exponential mapping. The evolution operator of the system, i.e.
the exponential of $L$ times the Hamiltonian is thus
\eqn\evol{e^{-H^{{\rm cyl},\tau} L }=
e^{{2\pi i
}(\tau (L_0-{c\over 24}) -\bar\tau(\bar L_0 -{c\over 24}))}\ .}

A convenient way to encode the information \hilbert\ is to look at
the partition function of the theory on a torus $\CT$. Up to a global
dilatation, irrelevant here, a torus may be defined
by its modular parameter $\tau$, $\Im m\,\tau>0$, 
such that its two periods are $1$ and $\tau$. Equivalently, 
it may be regarded as the quotient of the complex plane by the lattice
generated by the two numbers $1$ and $\tau$:
\eqn\torus{\CT= \IC/(\IZ \oplus \tau\IZ)\ , }
in the sense that points in the complex plane are 
identified according to $w\sim w'=w+ n+m \tau$, $n,m\in \IZ$. 
There is, however, a redundancy 
in this description of the torus:
the modular parameters $\tau$ and $M \tau$ describe the same torus,
for any modular  transformation $M\in PSL(2,\IZ)$.
The partition function of the theory on this torus is just the trace of the
evolution operator \evol, with the trace taking care of
the identification of the two ends of the cylinder into a torus
\eqn\torus{Z=\tr_{\CH_P} e^{2\pi i [\tau (L_0-{c\over 24}) -\bar\tau
(\bar L_0-{c\over 24})] } \ .}
Using \hilbert\ and the definition \chara\ of characters,
this trace may be written as
\eqn\toruspf{Z=\sum
N_{j\bar j} \, \chi_j(q) \chi_{\bar j}(\bar q)\qquad
q=e^{2\pi i\tau}\quad \bar q=e^{-2\pi i\bar\tau}
 . }
Let's stress that in these expressions, $\bar\tau$ is the complex conjugate of
$\tau$, and $\bar q$ that of $q$, and therefore, 
$Z=\sum N_{j\bj} \chi_j(q) \big(\chi_{\bj}(q)\big)^*$ 
is a {\it sesquilinear} form in the characters. 
 Finally this partition function must be intrinsically attached to the
torus, and thus be invariant under modular transformations. This
key observation, together with the expression of $Z$ as a sesquilinear
form in the characters, is due to Cardy \Camod. As we shall now show, it
opens a route to the classification of RCFTs.  
We have been led indeed to the~\dots
\smallskip
\nind{\bf Classification Problem} : {
\sl find all possible sesquilinear forms \toruspf\
with non negative integer coefficients that are modular invariant, and such
that $N_{11}=1$.}  \par\nind 
\smallskip
The extra condition $N_{11}=1$ expresses the unicity of the
identity representation (i.e. of the ``vacuum'').
\footnote{${}^{(3)}$}{Notice that the property of modular invariance is
just a necessary condition of consistency of the theory. It may be
--and it seems to happen-- that some modular invariants do not
correspond to any consistent CFT. The general conditions to be
fulfilled by a CFT to be fully consistent have been spelled out
by Sonoda \Sono\ and by Moore and Seiberg \MS. They amount essentially to
the consistency (duality equations) of the 4-point function
in the plane, and modular invariance (or covariance) of the 0- and 1-point
functions on the torus.
Nobody has been able, however, to analyse systematically these 
conditions besides the simplest cases of $sl(2)$-related theories. }
As explained in the previous section, the finite set of characters of any
RCFT, labelled by $\CI$, supports a unitary representation of the
modular group. This implies that any diagonal combination of
characters $Z=\sum_{i\in\CI} \chi_i(q) \chi_i(\bar q)$
is modular invariant. Are there other solutions to the problem?

The problem has been completely solved only in a few cases:
for the RCFTs with an affine algebra,
the $\slh(2)$ \CIZK\ and $\slh(3)$ \Ga\ theories at arbitrary level,
plus a host of cases with constraints on the level, e.g. 
the general $\slh(N)$ for $k=1$ \IDG, etc; 
associated coset theories have also been fully classified, including
all the minimal $c<1$ theories, $N=2$ ``minimal'' superconformal theories,
etc. A good review on the current
state of the art is provided by T. Gannon \Gannew.

\nind{\petit
In the case of CFTs with a current algebra, it is in fact better
to look at the same problem of modular invariants after replacing
in \toruspf\ all specialized characters by non-specialized ones, v.i.z.
$\sum N_{j\bar j} \chi_j(q,{\bf u}) \big(\chi_{\bj}(q,{\bf u})\big)^* $. 
Because these 
non-specialized characters are linearly independent, there is no ambiguity 
in the determination of the multiplicities $N_{j\bj}$ from $Z$.
This alternative form of the partition function may be seen as 
resulting from a modification of the energy-momentum tensor
$T(z)\to T(z) -{2\pi i\over L} ({\bf u},{\bf J}(z))  -{k\over 2} 
\({2\pi\over L}\)^2  ({\bf u},{\bf u})$, see \BPPZ. }

\subsec{The $\slh(2)$ cases}

\nind This was the first case fully solved.
With the notations introduced in the previous section for the
representations of $\slh(2)$,  the complete list
of modular invariant partition functions is as listed in Table 1.
A remarkable feature appears, namely an unexpected connection with
$ADE$ Dynkin diagrams. By this I mean that if we concentrate on the
{\it diagonal} terms of these expressions, their labels $\lambda$
turn out to be the Coxeter {\it exponents} of these Dynkin diagrams.
I recall that for such a diagram, the eigenvalues of its adjacency
matrix $G$ are of the form $2\cos {\pi \ell\over h}$,
$h$ the Coxeter number, and $\ell$ is an exponent taking
rank($G$) ($=$ number of vertices of $G$)
values between $1$ and $h-1$, with possible multiplicities.
Alternatively, the Cartan matrix $C=2 \II -G$ has eigenvalues
$4 \sin^2{\pi\ell\over 2 h}$. These Coxeter number and exponents are
listed in Table 2 (do not pay attention for the time being
to the last entry denoted $T_n$).
Anticipating a little on the following, let's notice that the
off-diagonal terms in the partition functions of Table 1 may also
be determined in terms of the data of the $ADE$ diagrams.

\vskip10mm
\bigskip
\settabs\+12345678901234567&890123456789012345678901234567890123456789012345
678901&\cr
\vbox{
\centerline{Table 1}
\noindent List of modular invariant partition functions in
terms of $SU(2)$ Kac--Moody characters $\chi_{\lambda}$. 
\smallskip
\hrule
\smallskip
\+ $ k\ge 0 $ & $ \displaystyle{\sum_{\lambda=1} ^{k+1} \vert
\chi_{\lambda}\vert^2} $ & $ (A_{k+1})$ \cr
\smallskip
\+ $ k=4\rho\ge 4  $ &
$ \displaystyle{\sum ^{2\rho-1}_{\lambda \> {\rm odd}\> = 1}
\vert\chi_{\lambda}+\chi_{4\rho+2-\lambda}\vert^2+2\vert\chi_{2\rho +1}
\vert^2} $ & $ (D_{2\rho+2}) $ \cr
\smallskip
\+ $ k=4\rho-2\ge 6  $ &
$ \displaystyle {\sum ^{4\rho-1} _{\lambda \> {\rm odd} \>=1}
\vert \chi_{\lambda}\vert^2
+ \vert \chi_{2\rho}\vert^2 + \sum _{\lambda \ {\rm even} \ =2}
^{2\rho-2} ( \chi_{\lambda} \chi _{4\rho-\lambda}^{\ast} +{\rm c.c.} ) }$
 & $(D_{2\rho+1})$ \cr
\smallskip
\+ $ k=10 $ & $ \vert \chi_1  + \chi_7 \vert^2+\vert \chi_4  + \chi_8
\vert^2 + \vert \chi_5  + \chi_{11} \vert^2 $ & $ (E_6)$ \cr
\smallskip
\smallskip
\+ $ k=16 $ & $ \vert \chi_1  + \chi_{17} \vert^2 + \vert \chi_5 + \chi_{13}
\vert^2 + \vert \chi_7  + \chi_{11} \vert^2  +\vert \chi_9 \vert^2
$ & $ (E_7)$ \cr
\+& $ \qquad +\lbrack ( \chi_3+\chi_{15} ) \chi_9 ^{\ast} +
{\rm c.c.}  \rbrack $ \cr
\smallskip
\smallskip
\+ $ k=28 $ & $ \vert \chi_1  +\chi_{11} + \chi_{19} + \chi_{29} \vert^2
 +\vert \chi_7  +\chi_{13} + \chi_{17} + \chi_{23} \vert^2
$ & $ (E_8) $ \cr }
\smallskip
\hrule
\bigskip
%
\bigskip
\bigskip
\vskip10mm

%
\vbox{
\centerline{Table 2: The graphs $A$-$D$-$E$-$T$, their Coxeter number and
their Coxeter exponents}
\medskip
\halign{ # & #  &  # & # \cr
 \quad & \qquad\qquad \qquad\qquad \qquad\qquad
&\qquad $h$  & \qquad{\rm exponents}\quad  \cr
\cr
 \qquad \qquad $A_n$& {\epsfxsize=3cm 
 	\epsfbox{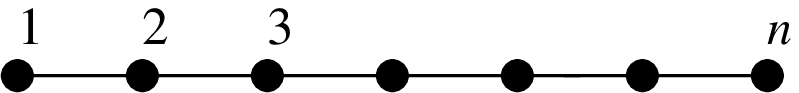}}  & \qquad   $n+1$  &\qquad $1,2,\cdots,n$ \cr
\cr
 \qquad \qquad $D_{\ell+2}$ &\raise -15pt\hbox{\epsfxsize=3cm 
 	\epsfbox{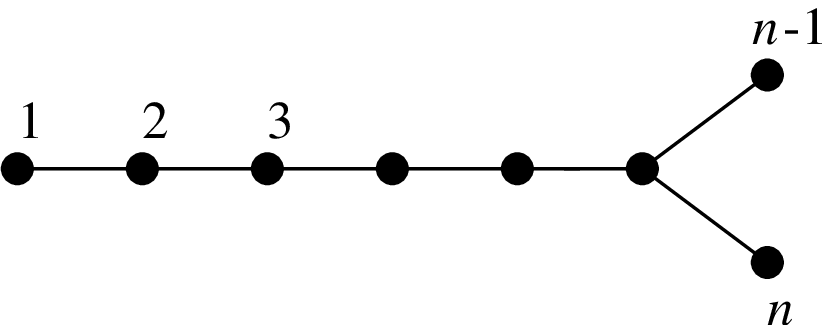}} &\qquad  $2(\ell+1)$     
	&\qquad $1,3,\cdots,2\ell+1,\ell+1$ \cr
 \qquad \qquad $E_6$ & \raise -5pt\hbox{\epsfxsize=25mm 
 	\epsfbox{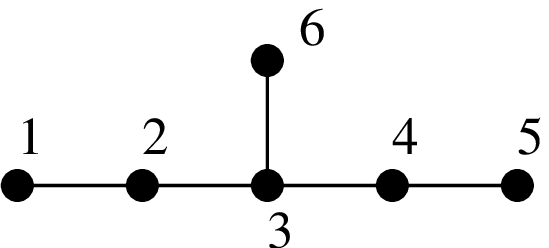}}  &\qquad 12  &\qquad  $1,4,5,7,8,11 $\cr
\cr
 \qquad \qquad $E_7$ &  \raise -5pt\hbox{\epsfxsize=3cm 
 	\epsfbox{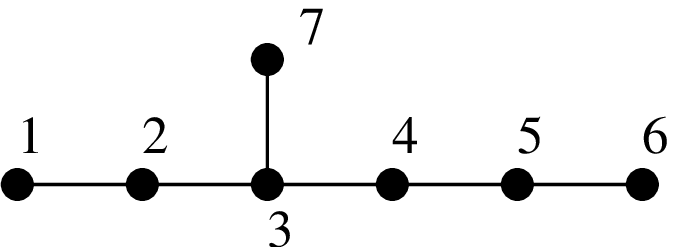}}&\qquad 18& \qquad $1,5,7,9,11,13,17$  \cr
\cr
 \qquad \qquad $E_8$ &
\raise -6pt\hbox{\epsfxsize=35mm 
	\epsfbox{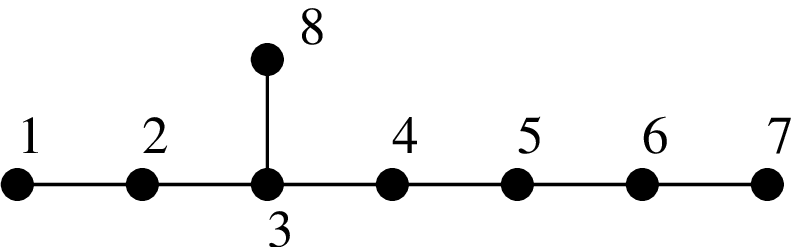}}&\qquad  30   & \qquad $1,7,11,13,17,19,23,29$ \cr
\cr
 \qquad \qquad $T_n$ & \raise -10pt\hbox{\epsfxsize=3cm 
 	\epsfbox{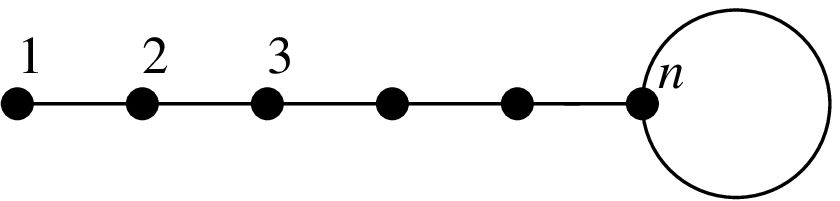}}&\qquad  $2n+1$ & \qquad $1,3,5,\cdots,2n-1$ \cr
}
\hrule}
\bigskip

%


\subsec{On ADE classifications}
\nind
The content of this section is not essential for what follows (except 
point (vii) below). 
The subject is so intriguing and so fascinating, however, that I cannot resist 
presenting  it. 

 It is well known that
there are many mathematical objects that fall in an $ADE$ classification
\Hazew.  The  list includes
\item{(i)} simple simply-laced
Lie algebras, i.e.  with roots of equal length \Bour;
\item{(ii)} finite reflection groups of cristallographic and of
simply-laced type \Coxhum;
\item{(iii)} finite subgroups of $SO(3)$ or of $SU(2)$, (or the associated
platonic solids);
\item{(iv)} Kleinian singularities \Slo;
\item{(v)} ``simple'' singularities, i.e. with no modulus \Arn;
\item{(vi)} finite type quivers  \Gab; 
\item{(vii)}  symmetric matrices with eigenvalues between $-2$ and
$+2$; 
\item{(viii)} algebraic solutions to the hypergeometric equation 
(\Hille, p 385);  
\item{(ix)} subfactors of finite index \Jo, see also D. Evans' lectures 
at this school;
\item{} and presumably others\dots 

\nind
To elaborate a little on these various items:
\item{(i)} 
Simple Lie algebras have been classified by Killing and Cartan;
restricting ourselves to the simply laced ones, i.e. with roots 
of equal length, leaves us with the $ADE$ cases.
\item{(ii)} Reflection groups are groups generated by reflections in
hyperplanes orthogonal to vectors $\{\alpha_a\}$ in the Euclidean
space $\IR^n$
called roots : $S_a~:~x\mapsto x-2 \alpha_a\,(\alpha_a,x)/(\alpha_a,\alpha_a)$. 
The group is of finite order iff the bilinear
form $( \alpha_a,\alpha_b)$ is positive definite.
This leads to a list $A_n,B_n\equiv C_n, D_n, E_6,E_7,E_8, F_4,G_2,
H_3,H_4,I_2(k)$, where the subscript gives the space dimension $n$ 
\Coxhum. 
If moreover, the condition that the root
system is crystallographic is imposed, (i.e. that for all pairs of
roots $\alpha,\beta$, $2 (\alpha,\beta)/(\beta,\beta) \in \IZ$),
only the cases $A$ to $G_2$ are left: they are
of course the Weyl groups of the Lie algebras of (i). Imposing
also the condition of simple lacedness leaves only $ADE$.
\item{(iii)} 
Finite subgroups of $SU(2)$ form two infinite series and three
exceptional cases: the cyclic groups $\CC_n$, the binary dihedral groups
$\CD_n$, the binary tetrahedral group $\CT$, the binary octahedral group $\CO$
and the binary icosahedral group $\CI$. 
It is natural to label them by $ADE$ as we shall see (Table 3).
A related classification is that of the 5 regular solids in 
three-dimensional  Euclidean space:
this may be the oldest $ADE$ classification, since it goes back to the
school of Plato; strictly speaking, only the exceptional cases
$E_6,E_7, E_8$ appear there, since $E_6$ is associated with the group
of the tetrahedron, $E_7$ with the group of the octahedron or of the
cube, and $E_8$ with  the group of the dodecahedron or of the icosahedron.
The cyclic and dihedral groups 
may be
thought of respectively  as the rotation invariance group of a pyramid 
and of a prismus of base a regular $n$-gon, but those are not
regular platonic solids.
\item{(iv)} Kleinian singularities:
 Let $\Gamma$ be a finite subgroup of $SU(2)$. It acts on
$(u,v)\in\IC^2$. The algebra of $\Gamma$-invariant polynomials in $u,v$
is generated by three polynomials $X,Y,Z$ subject to one relation
$W(X,Y,Z)=0$. The quotient variety $\IC^2/\Gamma$ is
parametrized by these polynomials $X,Y,Z$ and is thus 
embedded into the hypersurface $W(x,y,z)=0$, $x,y,z\in \IC^3$. This
variety $\CS$ is singular at the origin \Slo, see Table 3.
\item{(v)} Simple singularities, i.e.
polynomials $W(x_1,x_2, \cdots,x_p)$ with $\partial W/\partial x_j|_0=0$
for all $i=1,\cdots,p$, and with  no modulus (up to 
regular changes of the $x$), 
are also given by the same Kleinian polynomials $W$ (up to
the addition or subtraction of quadratic terms)  \Arn.
\item{(vi)} Quivers are oriented graphs, with vertices $a$ and edges
$e$. A representation of a quiver is the assignment to each
vertex $a$ of a non-negative integer $d_a$ and of 
 a vector space $V_a\equiv\IC^{d_a}$
and to each edge $e=(a\to b)$ of a linear map $f_e: \quad V_a\mapsto V_b$.
Two such representations $(V_a,f_e)$ and $(W_a,g_e)$ are equivalent
if there are linear maps $\varphi_a:\quad V_a\mapsto W_a$ such that
\Omit{the diagram
\def\hfl#1#2{\smash{\mathop{\hbox to 12mm{\rightarrowfill}}
\limits^{\scriptstyle#1}_{\scriptstyle#2}}}
\def\vfl#1#2{\llap{$\scriptstyle #1$}\left\downarrow
\vbox to 6mm{}\right.\rlap{$\scriptstyle #2$}}
\def\diagram#1{\def\normalbaselines{\baselineskip=0pt
\lineskip=10pt\lineskiplimit=1pt}  \matrix{#1}}
$$\diagram{
V_i &\hfl{\varphi_i}{}&W_i\cr
\vfl{f_e}{}&&\vfl{}{g_e}\cr
V_j &\hfl{\varphi_j}{}&W_j\cr
}$$
is commutative.}
$ \varphi_b \circ f_e=g_e\circ \varphi_a$
for all edges $e=(a,b)$.
 One proves that quivers of finite type, i.e. such that
they admit only a finite number of inequivalent 
indecomposable representations, are $ADE$ diagrams ! \Gab,\Slo.
\item{(vii)} Symmetric matrices with non negative integer entries
and eigenvalues between $-2$ and $2$ are the adjacency matrices
of the $ADET$ graphs of Table 2. The ``tadpoles'' $T_n=A_{2n}/\IZ_2$
may be ruled out if the condition of 2-colourability (or
``bipartiteness'') is imposed 
\GHJ. %
%
\bigskip
{ \halign{ # & # & # & # & # & # \cr
$\quad\qquad\Gamma$ & \qquad $\CC_n$ &\qquad\quad\  $\CD_n$ &\qquad 
$\CT$& \qquad $\CO$ & \qquad $\bar\CI$\cr
$\quad\qquad|\Gamma|$ & \qquad $n$ & \qquad\quad\  $4n$ & \qquad $24$ & \qquad $48$ &\qquad $120$ \cr
\quad\qquad$W$ & $X^{n} -Y Z$ & $X^{n+1}+ XY^2 +Z^2$ & $X^4+Y^3+Z^2$ & $X^3+X Y^3 +Z^2$ & $X^5+Y^3+Z^2$ \cr
\quad\qquad$G$ &\  $A_{n-1}$ &\  \  $D_{n+2}$ &\qquad  $E_6$ &\qquad $E_7$ &\qquad $E_8$ \cr
\cr}
Table 3: Finite subgroups of 
$SU(2)$, their orders, the Kleinian singularity and 
the associated Dynkin diagram. 
\medskip

The most obvious manifestation of the $ADE$ classification of these
objects is provided by the Dynkin diagram and its exponents.
In cases (i) and (ii) the Dynkin diagram encodes the geometry of the 
root system $\{\alpha_a\}$:
$(\alpha_a,\alpha_b)=C_{ab}$  the Cartan matrix
$=2 \II -G_{ab}$, $G$ the adjacency matrix, while the exponents
shifted by $ 1$  give the degrees of the invariant polynomials. 
Also in case (ii), the product over all the simple roots of the 
reflections $S_a$ defines the Coxeter element, unique up to
conjugation,  whose eigenvalues are $\exp 2i\pi m_i/h$, $m_i$
running over the exponents. 
In case (iii), as found by McKay \McK, things are subtle and beautiful: 
the corresponding {\it affine} Dynkin diagrams of $\hA$-$\hD$-$\hE$ type
describe the decomposition into irreducible representations of the
tensor products of the representations of $\Gamma$ by a two dimensional
representation. By removing the vertex corresponding to the identity
representation, one recovers the standard $ADE$ Dynkin diagrams
in accordance with Table 3. See
Appendix A for more elements on the McKay correspondence.
In the case (iv) of Kleinian singularities, one considers the 
{\it resolution} $\tilde{\CS}$ of the singular surface $\CS$: this is a 
smooth variety with a projection $\pi\,:\ {\tilde\CS}\to \CS$ which
is  one-to-one everywhere except above the singularity at $0$: one 
proves that the {\it exceptional divisor} $\pi^{-1}(0)$ is a connected 
union of spheres, 
$\pi^{-1}(0)=C_1 \cup \cdots  \cup C_r$, $C_i\cong \IP^1\IC$.
The Dynkin diagram of $ADE$ type listed on the last line of Table 3 
(or more precisely the negative of 
its Cartan matrix) describes the intersection form
of these  components $C_i$. 
In the case (v) of a simple singularity, one may consider its deformation
$W=\epsilon$ and look at the intersection of the homology cycles of its 
level set $\{x\in \IC_p,\ |x|< \delta\, |\ W(x)=\epsilon\}$, or at their
monodromy   as $\epsilon$ circles around the origin: the intersection 
is again encoded in the Dynkin diagram, while the monodromy is given 
by the Coxeter element of the associated Coxeter group.
Shifted by $-1$ the exponents give the degrees of the  homogeneous 
polynomials of the local ring $\IC[x_1,\cdots, x_p]/(\partial_{x_i} W)$   
of the simple singularity  \Arn\ etc, etc. 

This is just a sample of all the fascinating properties and 
crossrelations between these problems. 

In many cases, the classification follows from the spectral condition
(vii). In some others, however, the key point is the determination
of triplets of integers $(p,q,r)$ such that
${1\over p}+{1\over q}+{1\over r}>1$.
(Prove   that the $ADE$ list below includes all  the solutions except 
 $(p=1, q\ne r)$. )
Note  also that for the $D$ and $E$ cases, these integers
give the length of the three branches of the Dynkin diagram
 counted from the vertex of valency 3.
(The $A$ entry may look slightly artificial at this stage).
\par
\medskip
\indent \halign{ # & # & # & # & # & # \cr
\qquad\qquad$G$ & $A_{2n-1}$ & $D_{n+2}$ & $E_6$ & $E_7$ & $E_8$ \cr
\qquad\qquad $(p,q,r)$ & $(1,n,n)$ & $(2,2,n)$ & $(2,3,3)$ & $(2,3,4)$ & $(2,3,5)$
\cr}
{\petit\nind  The integers $(p,q,r)$ also appear in the definition 
of the binary polyhedral groups by generators and relations : 
 $R^p=S^q=T^q=RST$, $(RST)^2=1$. %
The relationship between these two occurrences
of $(p,q,r)$ is a nice manifestation of the ``dual McKay correspondence'' of 
\Bry.}

\medskip

To the above list, we have now added one more item:
the $\slh(2)$ modular invariant partition functions.
A natural question is: does this new item connect to any
previously known case? If it is so,  this may give
us a hint about what may be expected in other cases.
For example, is there some spectral property on  matrices
that would be related to the modular invariants of
$\slh(3)$ type? Or would the subgroups of $SU(3)$ be of relevance?
Or a certain class of singularities beyond the ``simple'' ones?

I think it is fair to say
that the question has not yet received a clear answer. We
do see connections between existing $ADE$ classifications and
some consequences of the $ADE$ classification of $\slh(2)$
modular invariants, but we do not see how they extend to
higher rank cases. Or at least, in no direct and systematic way \dots
This is what we shall show by discussing next the case of
$\slh(3)$ theories (sect. 2.3).
Then in  section 3,  we shall see that
the study of CFT in the presence of boundaries (``BCFT'')
gives us some new insight in these questions.
{\petit In the $\slh(2)$ case, there are two related classes of
CFTs for which another $ADE$ classification appears from another
standpoint: the $c<1$ (unitary) minimal models, and the ``simple''
$\CN=2$ superconformal field theories or their topological cousins
\WDVV.
Both may be obtained by the coset construction from
the $\slh(2)$ models, and inherit from them a variant of the $ADE$
classification. 
In the former case, it is known that $c<1$ minimal models admit a
lattice integrable realisation \VPun: the configuration space
of these lattice models
is the space of paths on a graph, and demanding that this space supports a
representation of the Temperley-Lieb algebra (a quantum deformation of the
symmetric group algebra and a known way to achieve integrability) and 
that the model is critical (in the sense of statistical mechanics) forces
us to restrict to graphs with eigenvalues of their adjacency matrix between
$-2$ and $2$, hence of $ADE$ type. 
On the other hand, the $\CN=2$ superCFTs have been argued
to admit a description of their ``chiral sector'' by a Landau-Ginsburg
superpotential which must be one of the simple singularities described
above, whence again of type $ADE$~\LVWM. Thus in these two cases, we have
an alternative way to see why and how $ADE$  appears.  Unfortunately, these
alternative standpoints are of little help in the higher rank cases:  the
class of graphs supporting a representation of the Hecke algebra associated
with $sl(N)$, $N>3$, is not yet known,   (for $N=3$, see 
\refs{\DFZun,\Soch,\Xu,\AO}) 
and the $\CN=2$ theories related to higher
$sl(N)$ are not all described by a Landau-Ginsburg potential.  }

{\petit There is a finer subdivision of items classified by $ADE$ into
two classes: those classified by $A,D_{{2\ell}}, E_6$, or $E_8$ and 
those classified by $D_{2\ell+1}$ or $E_7$. The distinction appears 
in many cases when one looks at positivity properties of some 
numbers, coefficients, etc. For example, in the list of Table 1, 
the modular invariants of the first class may be written as
sums of squares of linear combinations of $\chi$ with non negative
coefficients. I am not sure that the relevant positivity property
has been identified in all cases listed above (see \Zubkyoto\ 
for a discussion of some aspects of this issue). Ocneanu has shown 
that Dynkin diagrams of the first subclass admit a ``flat connection''
\OcnEK.
Another manifestation of the distinction is the existence or
non-existence of a fusion-like algebra, called the Ocneanu-Pasquier
algebra, attached to the Dynkin diagram \refs{\Pasq,\DFZdeu,\Zubkyoto}.
From the point of view of CFT, this apparently innocent looking 
distinction reveals a different structure of the theory.
The  block-diagonal modular invariant partition functions 
classified by $A,D_{2\ell},E_6,E_8$ may be regarded as diagonal
in terms of characters of some extended chiral algebra; the 
others are obtained from the latter by 
some twisting procedure \refs{\MS,\DV}.
All these considerations  extend beyond the case of $\slh(2)$ 
theories.}

\subsec{The $sl(3)$ case, the associated graphs}
\nind Let us turn to  the case of $\slh(3)$ for which complete 
results are now available. 
According to what was said above on 
the affine algebras ${\slh(N)}$ at a given level $k$, 
each integrable representation of $\slh(3)_k$ is labelled
by a weight $\lambda=\lambda_1 \Lambda_1+
\lambda_2 \Lambda_2$ subject to inequalities $\lambda_i \ge 1, 
\lambda_1+\lambda_2\le k+2$. With these notations, the complete 
list of modular invariants is given in Table 4 \Ga. It 
includes four infinite series and six exceptional cases.
In the same way as the modular invariants of $\slh(2)$ were 
associated with Dynkin diagrams, with the exponents of the 
latter giving the diagonal terms of the former,
 we find here the
\smallskip
\par\nind {\bf Fact} {\sl With each modular invariant $Z$ of 
$\slh(3)$ one may associate (at least) one graph, whose 
spectrum is encoded in the diagonal terms of $Z$. 
More precisely, the adjacency matrix of this graph has 
eigenvalues of the form: $S_{(2,1),\lambda}/S_{(1,1),\lambda}$, 
with a multiplicity equal to the diagonal element $N_{\lambda,\lambda}$
of $Z$. }

{\par\begingroup\parindent=0pt\leftskip=1cm\rightskip=1cm\parindent=0pt
\baselineskip=11pt
\midinsert
\epsfxsize=15cm
\centerline{\epsfbox{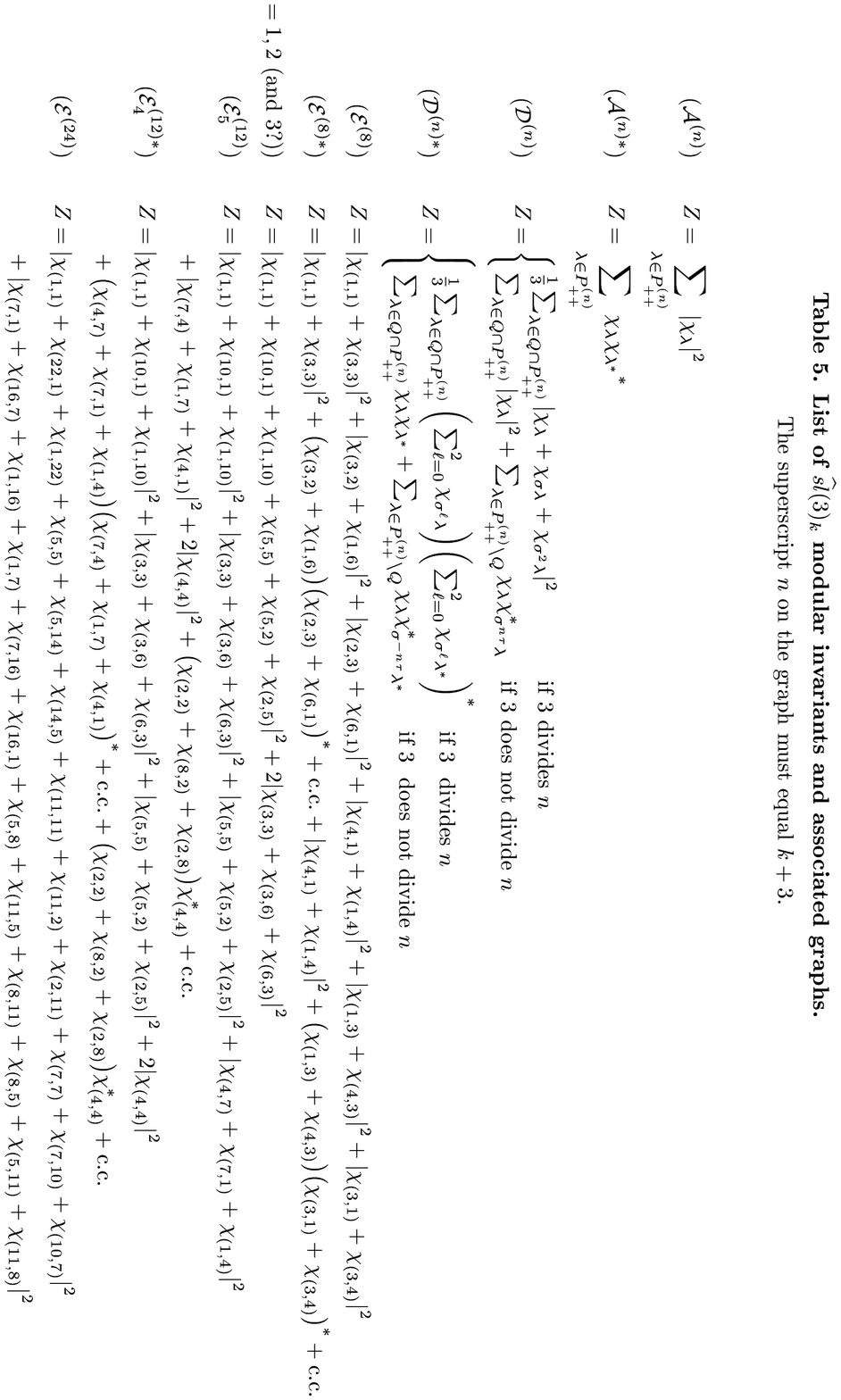}}
\vskip 12pt
\endinsert\endgroup\par
}

This fact was originally proposed as an  Ansatz  and 
graphs were found by empirical methods \DFZun. Later, more systematic 
techniques to determine the graph were developed in a variety of cases
\PZdeu, but the physical interpretation of the graph itself 
remained unclear until recently, when it took a   
new perspective in the light of boundary conformal 
field theory (Lecture 3). On a more abstract level, this  also inspired 
developments by Ocneanu, by Xu and by B\"ockenhauer, Evans and Kawahigashi 
\refs{\Ocn,\Xu,\BEK}. The list of $\slh(3)$ graphs displayed in 
Fig. 1 and 2 were recapitulated in a recent work with Behrend, Pearce and 
Petkova \BPPZ. 

\fig{The graphs of $\slh(3)$}{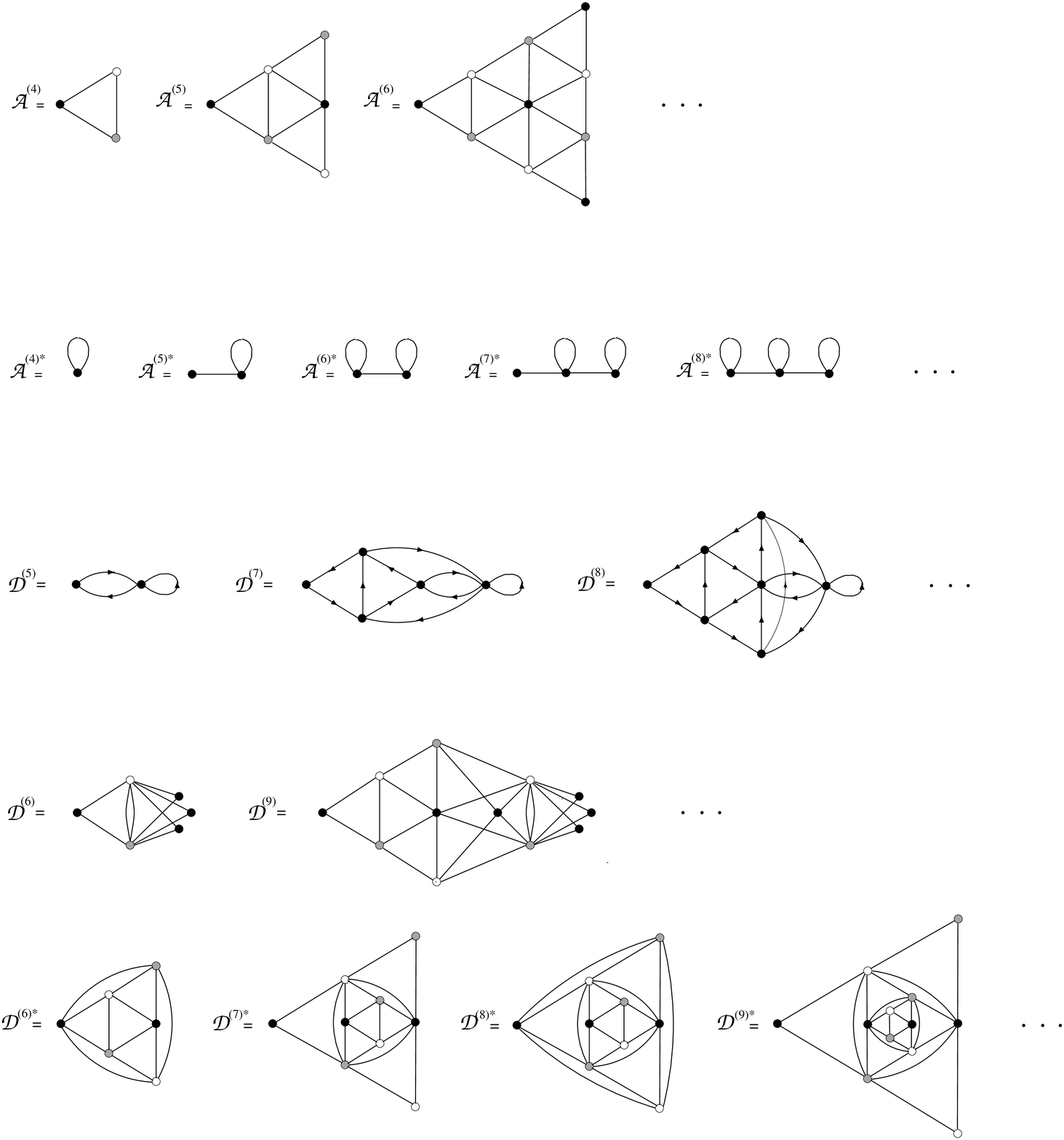}{165mm}\figlabel\sutroisun
\fig{The graphs of $\slh(3)$ (continued)}{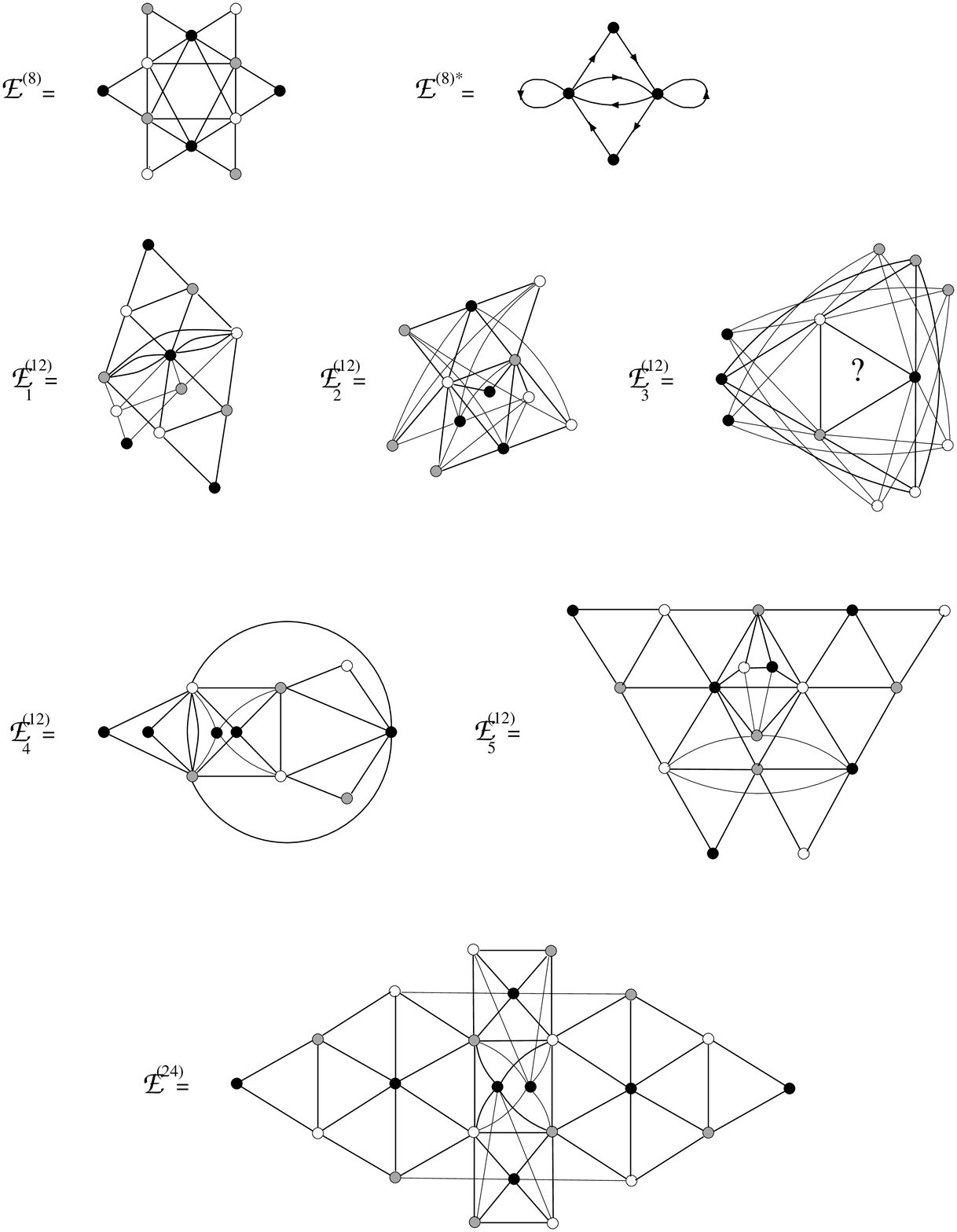}{135mm}\figlabel\sutroisdeu

\bigskip
{\petit Let us discuss briefly the properties 
of these graphs (Fig. 1 and 2).\par\noindent
$\star$ Most of them turn out to be 3-colourable:  a colour,  black ($b$), 
gray ($g$) or white ($w$), 
may be assigned to each vertex, and it is understood
that the oriented edges go as: $ b \to g\to w \to b$. Some of the graphs, 
however, are not 3-colourable, and then the orientation of the edges 
has been explicitly displayed whenever necessary, while the absence of
arrow in these cases means that the edge carries both orientations.\par\nind
$\star$ All the graphs listed in Fig. 1 and 2 satisfy the spectral
property stated above, but many others also do, which are not
associated with any modular invariant \DFZun. 
\par\nind
$\star$ There are a few cases for which two graphs are
associated with the same modular invariant, for example
$\CD^{(6)}$ and $\CD^{(6)*}$, or $\CE^{(12)}_1$ and $\CE^{(12)}_2$.
The graph  $\CE^{(12)}_3$ which is isospectral with the two latter 
and was believed so far to be also associated with the same modular
invariant, is discarded  by A. Ocneanu on the basis 
that it does not support a system of ``triangular cells'' \AO,
i.e., presumably, that the corresponding CFT has no consistent 
Operator Algebra. It is  thus marked with a question mark on Fig. 2.
\par\nind
$\star$ Many of these graphs may be obtained by the following  
procedure, inspired by the McKay correspondence for $SU(2)$. 
Given a finite subgroup $\Gamma$ of $SU(3)$, the decomposition 
into irreducibles $(b)$ of the tensor product of each irreducible 
representation $(a)$ of $\Gamma$ by the restriction to $\Gamma$ of the
defining  three-dimensional representation of $SU(3)$ yields a 
matrix $\hat G$: $ (a) \otimes (3) =\oplus_b {\hat G}_{ab} (b)$. 
Some appropriate truncation of the graph of $\hat G$ may then yield
a graph adequate for our problem. Contrary to the case of $SU(2)$, 
however, things are neither systematic --which vertices/edges have to 
be deleted is not clear a priori-- nor exhaustive: some graphs like 
${\cal E}^{(24)}$ in Fig. 2 are not reached by this procedure.
\par\nind
$\star$ It may be interesting, in view of point (v) of the $ADE$ list above,
to note that some of these graphs enable one to construct a reflection 
group associated with a singularity \Ztani,\GZV. 
\par\nind
%
$\star$ For a proof that the list of graphs is complete
from the subfactor perspective see \AO. See also the discussion 
in \Xu,\BEK, where several of these graphs have been reproduced.    
}

\medskip
Of course, there is nothing special with $\slh(3)$ at this stage, 
and everything could be repeated for higher rank, 
except that no complete list of modular invariants 
nor of graphs is known in these cases.  Also, for $\slh(N)$, it is
in fact a collection of $N-1$ graphs, labelled by the fundamental
representations of $SU(N)$,  which must be provided. Complex 
conjugate representations give graphs with all orientations reversed, 
and the first $[N/2]$ are thus sufficient. 

\subsec{General case}
\nind In general, we expect that a graph (or a collection of graphs) 
will be associated with  
any ``rational'' conformal field theory, (see section 1.4), 
with the spectrum 
of its adjacency matrix determined by the diagonal terms 
of the  partition function. Among the pairs $(j,\bj)$ appearing 
in $Z$, a special role will therefore be played by the diagonal subset 
\eqn\expo{\calE=\{j| j=\bar j, N_{jj}\ne 0\} \ ,}
the elements of which, the ``exponents'' of the theory, will be 
counted with the multiplicity $N_{jj}$.
Note that $\CE$ is stable under conjugation: $j$ and $j^*$ occur 
with the same multiplicity. 
The justification of this association of one (or several) 
graph(s) with a CFT will appear in the next section. 

\newsec{Boundary Conformal Field Theory}
\subsec{RCFT in the half-plane, boundary conditions and operator content}
\nind
We now turn to the study of RCFTs in a half-plane. There are
several physical reasons to look at this problem, --critical systems
in  the presence of a boundary, open strings and generalized D-branes,
one-dimensional electronic systems with  ``quantum impurities'' etc.
Here we shall only look at the new information and perspective that
this situation gives us in the classification problem of RCFT.

In a half-plane, the admissible diffeomorphisms must respect the
boundary, taken as the real axis: thus only real analytic changes
of coordinates, satisfying $\epsilon(z)=\bar\epsilon(\bz)$ for $z=\bz$ real,
are allowed. The energy momentum itself has this property:
\eqn\bcT{T(z)=\bar T(\bar z)|_{\rm{real\ axis}}\ , }
which expresses simply the
absence of momentum flow across the boundary
and which enables one to extend the  definition of $T$
to the lower half-plane by $T(z):=\bar T(z)$ for $\Im m\, z<0$.
There is thus only one copy of the Virasoro algebra $L_n=\bar L_n$.
This continuity equation \bcT\ on  $T$  extends
to more general chiral algebras and their generators, at the price
however of some complication. In general, the continuity
equation on generators of the chiral algebra involves some
automorphism of that algebra:
\eqn\bcW{ W(z)=\Omega\bar W(\bar z)|_{\rm{real\ axis}}}
(see \BPPZ\ and further references therein).

The half-plane, punctured at the origin, (which introduces a distinction
between the two halves of the real axis), may also be conformally mapped
on an infinite horizontal strip of width $L$ by $w={L\over \pi} \log z$.
Boundary conditions, loosely specified at this stage by labels $a$ and
$b$, are assigned to fields on the two boundaries $z$~real~$> 0$, $< 0$
or $\Im m\, w=0, L$.
For given boundary conditions on the generators of the algebra
and on the other fields of the theory, i.e. for given
automorphisms $\Omega$ and given $a,b$,
we may again use a description of the system
by a Hilbert space of states $\calH_{ba}$ (we drop the dependence on $\Omega$
for simplicity). On the half-plane or on the finite-width strip,
only {\bf one copy} of the Virasoro algebra, or of the chiral algebra $\CA$
under consideration, acts on $\CH_{ba}$, and
this space decomposes on representations of Vir or $\CA$ according to
\eqn\Hab{ \calH_{ba}= \oplus n_{ib}{}^a \calV_i\ ,}
with  a new set of multiplicities $ n_{ib}{}^a\in{\IN}$.
The natural Hamiltonian on the strip is 
the translation operator in $\Re e\, w$, hence, mapped back in the half-plane
\eqn\Hstrip{ H_{b|a}={2\pi\over L}\left(L_0-{c\over 24}\right)\ . }

To summarize, in order to fully specify the operator content of the theory
in various configurations, we need not only determine the multiplicities
``in the bulk''$ N_{j\bar j}$ of \hilbert, but also the possible
boundary conditions $a,b$ on a half-plane and the associated multiplicities
$n_{ib}{}^a $. This will be our task in the following, and as we shall
see, a surprising fact is that the latter have some bearing on the former.

\fig
{ The same domain seen in different coordinates: a semi-circular 
annulus, with the two half-circles identified, a rectangular 
domain with two opposite sides identified, and a circular annulus. }
{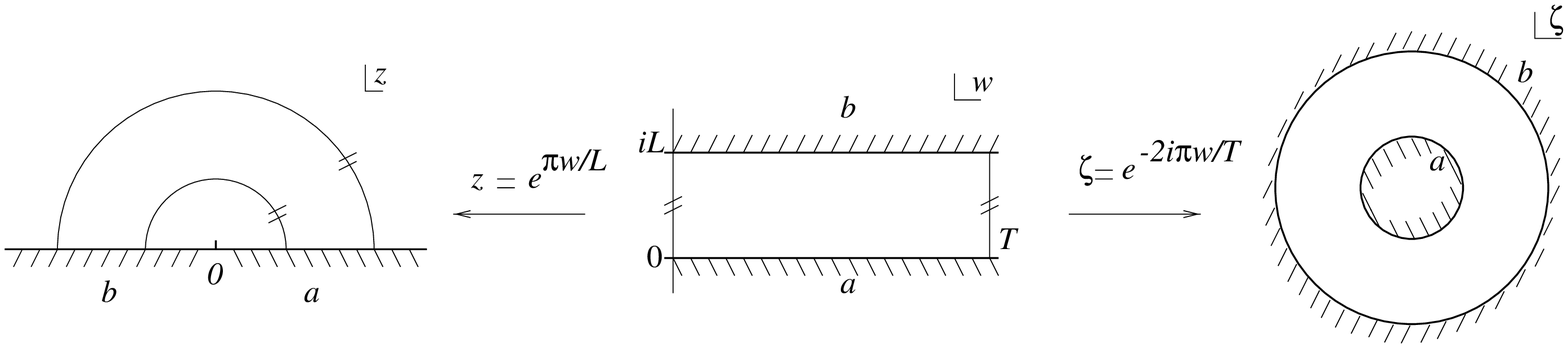}{17truecm}\figlabel\annulus

\subsec{Boundary states}
\nind In the same way that we found useful to chop a finite segment of the
infinite cylinder and identify its ends to make a torus, it is
suggested to consider a finite segment of the strip -- or a
semi-annular domain in the half-plane-- and identify its edges,
thus making a cylinder. 
This cylinder can be mapped back into an annular domain in the plane,
with open boundaries.
More explicitly, consider the segment $0\le \Re e\,w\le T$ of the strip
--i.e. the semi-annular domain in the upper half-plane
comprised between the semi-circles of radii 1 and $e^{\pi T/L}$, the latter
being identified. It may be
conformally mapped into an annulus in the complex plane
by $\zeta=\exp(- 2i\pi w/T)$, of radii $1$ and $e^{2\pi L/T}$, see Fig. 3. By
working out the effect of this change of coordinates on the energy-momentum
$T$, using \varT, one finds that \bcT\ implies
\eqn\newbcT{ \zeta^2 T(\zeta)= \overline{\zeta}^2 \overline{T}(\bar \zeta)
\quad{\rm for}\ |\zeta|=1,\ e^{2\pi{L\over T}}\ .}
{\petit \Ex : assuming that $W$ transforms as a primary field of
conformal weight $(h_W,0)$, find the corresponding condition on
$W(\zeta)$.}\par\nind
After radial quantization, this translates into
 a condition on {\it boundary states} $|a\rangle\, ,\ |b\rangle\, \in \CH_P$
which describe the system on these two boundaries. 
\eqn\bcsta{(L_n-\bar L_{-n})|a\rangle=0 }
(and likewise for $|b\ket$). Analogously
 $(W_n- (-1)^{h_W} \Omega(\bar W_{-n}))|a\rangle=0$, with $h_W=$ spin of $W$.

\medskip
We shall now look for a basis of states, solutions of this linear
system of boundary conditions.
One may look for solutions of these equations in each
$\calV_j\otimes \calV_{\bar j} \subset \CH_P$,
since these spaces are invariant under the action of the two copies
of Vir or of the chiral algebra $\CA$.  Consider only for simplicity
the case of the Virasoro generators.

\smallskip
\nind {\bf Lemma}\ {\sl
There is an independent ``Ishibashi state'' $|j\rrangle$, 
solution of \bcsta,  
for each $j=\bar j$, i.e. $j\in \calE$, the set of exponents.}
\smallskip\nind
Proof (G. Watts): Use the identification between states
$ |a\ket \in \calV_j\otimes \calV_{\bar j}$  
and operators $X_a\in \hbox{Hom}(\calV_{\bar j}, \calV_{j})$,
namely
 $|a\ket=\sum_{n,\bar n} a_{n,\bar n} |j,n\ket \otimes |\bar j, \bar n\ket $
$\leftrightarrow 
X_a =\sum_{n,\bar n} a_{n,\bar n} |j,n\ket \bra\bar j, \bar n| $.
Here we make use of the scalar product in $\calV_{\jb}$ for
which $\bar L_{-n}  =\bar L_n^\dagger $, hence 
\bcsta\ means that $L_n X_a=X_a L_n$, i.e.
$X_a$ intertwines the action of Vir in the two irreps
$\calV_j$ and $ \calV_{\bar j}$. By Schur's lemma, 
this implies that they are 
equivalent,  $\calV_j \sim \calV_{\bar j}$, i.e. that their labels
coincide $j=\bar j$ and that $X_a$ is proportional to $P_j$, the projector
in $\CV_j$. We shall denote $|j\rrangle$ the corresponding state,
solution to \bcsta. \par\nind
{\petit Since ``exponents'' $j\in \Exp$ may have some
multiplicity, an extra label should be appended to our
notation $|j\rrangle$. We omit it for the sake of simplicity.
The previous considerations extend with only notational complications
to more general chiral algebras and their possible
gluing automorphisms $\Omega$. See \BPPZ\ for more details on
these points. Also, in this discussion, I have been a bit cavalier on 
some points: the fact that these Ishibashi
states have no finite norm and thus do not really belong to $\CH$, 
and the use of Schur's lemma in this context
would require some justification:  See \FS\ for an alternative  and more 
precise discussion.  }
\bigskip

The normalization of this ``Ishibashi state'' requires some care.
One first notices that, for $\tilde q$ a real number between 0 and 1,
\eqn\chIsh{\llangle j'| \tilde q^{\oh(L_0+\bar L_0 -{c\over 12})}|j\rrangle
=\delta_{jj'}{\chi_j(\tilde q)}}
up to a constant that we choose equal to 1.
It would seem natural to then define the norm of these states by
the limit $\tilde q\to 1$ of \chIsh. This limit diverges, however, and
the adequate definition is rather
\eqn\normIsh{\llangle j|\!|j'\rrangle =
\delta_{jj'} {S_{1j}}}
{\petit This comes about in the following way: a natural
regularization of the above limit is:
\eqn\newnorm{\llangle j|\!|j'\rrangle =
\hbox{lim}_{\tilde q\to 1}
q^{c/24} \llangle j'| \tilde q^{\oh(L_0+\bar L_0 -{c\over 12})}
|j\rrangle}
where $q$ is the modular transform of $\tilde q=e^{-2\pi i/\tau}$,
$q=e^{2\pi i\tau}$. In a (``unitary'') theory in which the identity
representation (denoted 1) is the one with the smallest conformal weight,
show that in the limit $q\to 0$, the r.h.s. of  \newnorm\ reduces
to \normIsh. In non unitary theories, this limiting procedure fails, 
but we keep \normIsh\ as a definition of the new norm.}

At the term of this study, we have found a basis of solutions to
the constraint \bcsta\ on boundary states, and it is thus legitimate
to expand the two states attached to the two boundaries of our domain
as
\eqn\expbs{ |a\rangle =\sum_{j\in \calE} {\psi_a^j \over \sqrt{S_{1j}}}
\;|j\rrangle }
with coefficients denoted $\psi_a^j$, and likewise for 
$|b\rangle$. We 
define an involution $a\to a^*$ 
on the boundary states 
by $\psi_{a^*}^j=\psi_a^{j^*}=(\psi_a^{j})^*$, (recall that $j\to j^*$ is an
involution in $\calE$). One may show \RS\ that it
is natural to write for the conjugate state
\eqn\expconjs{\langle b| =\sum_{j\in \calE}  \llangle  j|\,
{\psi_{b^*}^j \over \sqrt{S_{1j}}}\ .}
 As a  consequence
\eqn\scalpr{
\bra b\|a\ket
=\sum_{j\in \calE}{\psi_a^j\left(\psi_b^j\right)^*\over S_{1j}}
\llangle j \| j\rrangle =
\sum_{j\in \calE}
{\psi_a^j \left(\psi_b^j\right)^*} }
so that  the orthonormality of the boundary states is equivalent  to
that of the $\psi$'s.

\subsec{Cardy equation}
\nind Let us return 
to the annulus $ 1\le |\zeta|\le e^{2\pi{L/ T}}$
considered in last subsection, or equivalently 
to the cylinder of length $L$ and perimeter $T$, 
with boundary conditions (b.c.)
$a$ and $b$ on its two ends. Following Cardy \Cabc, we shall
compute its partition function $Z_{b|a}$ in two different ways.
If we regard it as resulting from the evolution between the
boundary states $|a\ket$ and $\bra b|$, with $\tilde q^\oh=e^{-2\pi{L/T}}$, 
we find%
\eqn\Zabun{\eqalign{
Z_{\b|\a}&= \langle
\b| ({\tilde q }^{\oh(L_0+\bar L_0 -{c\over 12})}
|\a\rangle               
=\sum_{j,j'\in\calE}
{\left(\psi_b^j\right)\!{}^*\,\psi_a^{j'} \over S_{1j}}
\llangle j|\tilde{q}^{\,{1\over 2}(L_0+\overline{L}_0-{c\over 12})}
|j'\rrangle\cr
&= \sum_{j\in\calE}
\psi_a^j \left(\psi_b^j\right)\!{}^*\, {\chi_j(\tilde{q})\over S_{1j}}\ .\cr
} }

\fig
{Two  alternative computations of the partition
function $Z_{b|a}$: (a) on the cylinder, between the
boundary states $|a\rangle$ and $\langle b|$,  (b) as a periodic time
evolution on the strip, with boundary conditions $a$ and $b$.}{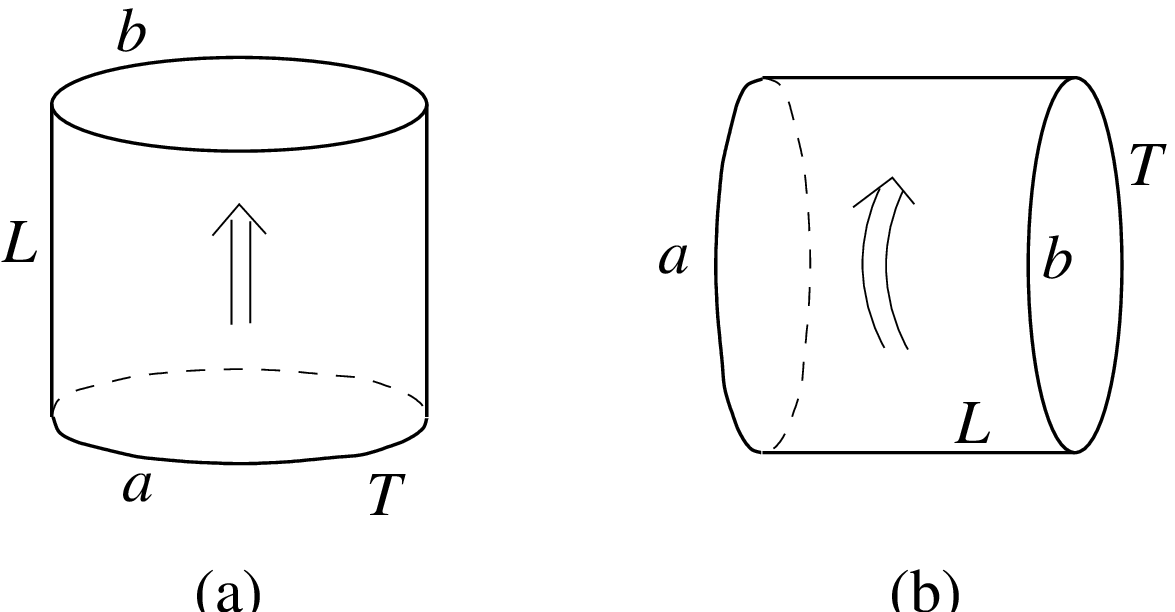}
{8cm}\figlabel\cardyeq

On the other hand, if we regard it as resulting from the
periodic  ``time''
evolution on the strip with b.c. $a$ and $b$, using
the decomposition \Hab\ of the Hilbert space $\CH_{ba}$,
and with $q=e^{-\pi T/L}$
\eqn\Zabde{
Z_{b|a}(q)=\sum_{i\in\calI} n_{ib}{}^a \chi_i(q)\ .}
(Note that string theorists would refer to these two situations as 
(a): the tree approximation of the propagation of a closed string; 
(b) the one-loop evolution of an open string). 
Performing a modular transformation on the characters
$\chi_j(\tilde q)=\sum_i S_{ji^*}\chi_i(q)$ in \Zabun,
and identifying the coefficients of $\chi_i$ yields
\eqn\cardy{
n_{ia}{}^b =\sum_{j\in\calE} \, {S_{ij}\over S_{1j}}\,
\psi_a^j \left(\psi_b^j\right)\!{}^*  \ ,}
a fundamental equation for our discussion that we refer to as Cardy 
equation \Cabc.
In deriving this form of Cardy equation, we have made use
of the first of the following symmetry properties of $n_{ia}{}^b$
\eqn\symnab{
n_{i\a}{}^\b =n_{i^*\b}{}^{\a} 
\ }
which follow from the previous relations on $\psi$ and of the symmetries
of $S$.
\nind{\petit (Comment: this identification of coefficients of specialized
 characters is in general not justified, as the $\chi_i(q)$ are not linearly
independent. As in sect. 2, it is better to 
generalize the previous discussion, in a way which introduces 
non-specialized --and linearly independent-- characters. This has
been done in \BPPZ\ for the case of CFTs with a current algebra.
Unfortunately, little is known about other chiral algebras and their
non-specialized characters.)}
\medskip

Let us stress that in \cardy, the summation runs over $j\in \CE$, i.e.
this equation incorporates some information on the spectrum of the 
theory ``in the bulk'', i.e. on the modular invariant partition function 
\toruspf.
 
Cardy equation \cardy\ is a non linear constraint relating
a priori unknown complex coefficients $\psi^j$ to integer multiplicities
$n_{ia}{}^b$. We need additional assumptions to exploit it.

We shall thus assume that
\item{$\bullet$} we have found an orthonormal set  of boundary states
$|a\rangle$, i.e. satisfying
\eqn\ortho
{ (n_1)_a{}^b=\sum_{j\in\calE}\psi_a{}^j(\psi_b{}^j)^*=
\delta_{ab}\ ;}
\item{$\bullet$} we have been able to construct a  {\it complete} set
 of such boundary states $|a\rangle$
\eqn\compl{ \sum_a \psi_a{}^j(\psi_a{}^{j'})^*=\delta_{jj'} \ .}
Note that the second assumption --which looks far from obvious to me--
implies that
$$ \#\ \hbox{boundary states}=\#\ \hbox{independent Ishibashi states}=|\calE|\ .$$

\subsec{Representations of the fusion algebra and graphs}
\nind
Return to Cardy equation \cardy\ and observe that it gives a
decomposition of the matrices $n_i$, defined by $(n_i)_a{}^b=n_{ia}{}^b$,
into their orthonormal eigenvectors $\psi$ and their eigenvalues
$S_{ij}/S_{1j}$. Observe also that as a consequence of Verlinde formula 
\verl, these eigenvalues form a one-dimensional representation of
the fusion algebra
\eqn\onedrep{
{S_{i\ell}\over S_{1\ell}}\; {S_{j\ell}\over S_{1\ell}}
=\sum_{k\in\calI} \calN_{ij}{}^k\; {S_{k\ell}\over S_{1\ell}}\, , 
\qquad \forall i,j,\ell\in \CI\ .} 
Hence the matrices $n_i$ also form a representation
of the fusion algebra
\eqn\nfus
{n_i\;n_j=\sum_{k\in\calI} \calN_{ij}{}^k\,n_k \ }
and they thus commute. Moreover, as we have seen above, 
they satisfy $n_1=I$, $n_i^T=n_{i^*}$.

Conversely, consider any ${\IN}$-valued matrix representation of 
the Verlinde fusion
algebra $n_i$, such that $n_i^T=n_{i^*}$. Since the algebra is
commutative, $[n_i,n_i^T]=[n_i,n_{i^*}]=0$. The $\{n_i\}$ form
 a set of {\it normal }
matrices, hence are diagonalizable in a common orthonormal basis. Their
eigenvalues are known to be of the form $S_{ij}/S_{1j}$. They may thus 
be written as in \cardy. Thus any such ${\IN}$-valued matrix 
representation of the Verlinde fusion algebra
 gives a (complete orthonormal) solution to Cardy's equation.

\medskip\nind 
{\bf Conclusion:}\
\medskip

\encadre{
$${\IN}\hbox{-valued matrix representation  of the fusion algebra}$$
$$\Longleftrightarrow \hbox{ \quad Complete, orthonormal solution of Cardy equation}$$
}

\nind
Moreover, since ${\IN}$-valued matrices are naturally interpreted as graph
adjacency matrices,  graphs appear naturally!

{\petit The relevance of the fusion algebra in the solution of Cardy equation
had been pointed out by Cardy himself for diagonal theories \Cabc\ 
and foreseen  
in general in \DFZun\ with no good justification; the importance of the
assumption of completeness of boundary conditions
 was first stressed by Pradisi et al \PSS.}

\subsec{The case of $\widehat{sl}(2)$ WZW theories}
\noindent
{\bf Problem:}\
{\sl Classify all ${\IN}$-valued matrix reps of $\widehat{s\ell}(2)_k$ fusion
algebra with $k$ fixed.}

\noindent The algebra is generated recursively by $n_2$
\eqn\recurr{n_1=I, \qquad n_2\;n_i=n_{i+1}+n_{i-1},\quad i=2,\ldots,k}
$$
\hbox{$S$ real}\ \Rightarrow\  n_i=n_i^T \ .
$$

\noindent
Even though $\psi^j$ and $\calE$ are yet unknown we know from \Ssld\ 
that $n_2$ has eigenvalues of the form
\eqn\evalu{
\gamma_j={S_{2j}\over S_{1j}}=2\cos{\pi j\over k+2},\quad j\in\calE \ .
}
But as discussed already in sect. 2.2, ${\IN}$-valued matrices $G$
with spectrum $|\gamma|<2$ have been classified. They are
the \hbox{adjacency} matrices  either of the $\hbox{$A$-$D$-$E$}$ Dynkin
$\hbox{diagrams}$ or of the ``tadpoles'' $T_n={A_{2n}/{\IZ}_2}$. 
Thus as a consequence of equation \cardy\ alone, 
 for a $\slh(2)$ theory at level $k$, the possible boundary conditions
are in one-to-one correspondence with the vertices of one of 
these diagrams $G$, with Coxeter number $h=k+2$. If we remember, however,
 that the set $\cal E$  must appear in one of the modular invariant torus
partition functions, the case $G=T_n$ has to discarded, and we are 
left with $ADE$. (Up to this last step, this looks like the 
simplest route leading to the $ADE$ classification of $\slh(2)$ theories.)
We thus conclude that for each $\slh(2)$ theory classified by a 
Dynkin diagram $G$ of $ADE$ type
\eqnn\conclu
$$\eqalign{ \calE &=\hbox{Exp}(G),\qquad \hbox{dim}(n_i)=|\calE|=|G| \cr
\hbox{complete } & \hbox{ orthonormal b. c. } = a,b,\cdots\ :\ 
\hbox{vertices of $G$} \cr
n_2&=\hbox{ adjacency matrix of $G$ }\cr
n_i&=\hbox{``$i$-th fused adjacency matrix'' of $G$ }\cr
\psi^j &=\hbox{eigenvector of $n_2$ with eigenvalue $\gamma_j$} \cr
}
$$

One checks indeed that 
the matrices $n_i$, given by equation \cardy, together with 
\Ssld, have only non negative integer elements. Let us pause 
a little to examine the remarkable properties of these matrices
which seem to play an ubiquitous role\dots

\subsec{Side remark: the $\slh(2)$ ``intertwiners''}
\nind
Let $G$ be a given Dynkin diagram of $ADE$ type, with Coxeter number $h$.
We want to look more closely at properties of the matrices $n_i$
just defined. Explicitly, using \Ssld, 
\eqn\intew{n_{ia}{}^b=\sum_{\ell\in \Exp(G)}
{\sin{\pi \ell i \over h}\over \sin{\pi \ell\over h}}
\psi_a^\ell \psi_b^{\ell *}\ .}
As stated above, all their matrix elements are non negative integers, 
and they form a representation of the fusion algebra. But also, 
regarded as rectangular matrices for fixed $a$, they satisfy
an intertwining property $A n =n G$ 
with the adjacency matrix $A$ of the Dynkin 
diagram $A_{h-1}$. More explicitly
\eqn\intww{ \sum_{j\in A_{h-1}} A_{ij} n_{ja}{}^b=
\sum_{c\in G} n_{ia}{}^c G_{cb}}
These matrices $n$ have made repeated appearances in various 
contexts.
\item{(1)}{In CFT: } For an appropriate choice of a subset 
$T$ of the vertices of $G$ 
\footnote{${}^{(4)}$}{the set of ambichiral vertices in the 
language of A. Ocneanu}  and in particular of a special
vertex denoted 1,  $1\in T$, one may write the torus
partition functions of type $A, D_{{\rm even}}, E_6, E_8$
 in the form
\eqn\Ztor{ Z_{{\rm torus}}= \sum_{\a\in T} |\hat\chi_a|^2}
where $\hat\chi_a:= \sum_i n_{i1}{}^a \chi_i$ are combinations
of characters of the original algebra, interpreted as characters
of a larger ``extended''   algebra (see sect. 2.2 {\it in fine}). 
This formula, originally found
empirically in \DFZun\ and justified later in a  variety of
cases in \PZdeu, has been extended to the missing cases
$D_{{\rm 2\ell+1}}, E_7$ by use of a relative twist between the 
right and left characters $\hat\chi$ pertaining to the $A_{4\ell-1}$ 
and $D_{10}$ cases, respectively \Ocn, in agreement with the general
result of \refs{\DV,\MS} recalled in sect 2.2 {\it in fine}. 
The general formula is thus 
\eqn\block{Z_{{\rm torus}}= \sum_{\a\in T} \hat\chi_a(q)
\left(\hat\chi_{\zeta(a)}(q)\right)^* \ .}
Here $\zeta$ is an automorphism of the fusion algebra of the representations
of the extended algebra labelled by $a\in T$.
This formula has also received a new interpretation in the light 
of the work of \Xu\BEK: see D. Evans'  lectures at this school, 
and compare his expression $\langle\alpha^+_i, \alpha^-_j\rangle$
for the matrix $N_{ij}$, with that coming from  \block\ 
$N_{ij}=\sum_{a\in T} n_{a1}{}^i \,n_{\zeta(a)1}{}^j$. 
\item{(2)}{Lattice models, graphs, operator algebras}. \
These same matrices $n$ 
also give the decomposition of representations of the Temperley-Lieb 
algebra on the space of paths from $a$ to $b$ on the graph $G$
onto the irreducible ones given by  the paths from 1 to $i$ on graph
$A_{h-1}$ \PaSa \
$$R^{(G)}_a{}^b =\oplus_i n_{ia}{}^b R^{(A)}_1{}^i$$
Another manifestation of this is that they give the 
counting of ``essential'' paths \Ocn. See Appendix B for details.
\item{(3)} Kostant polynomials in McKay's correspondence: More
surprisingly, maybe, these matrices also appear in the explicit 
expressions of the so-called Kostant polynomials, in the context of McKay
correspondence, \Kos: see Appendix A. In that context too, 
the matrices $n$ have received a very neat group
theoretical interpretation by Dorey, in terms of the action of the Coxeter 
element on simple roots \Dor.
\item{(4)} Finally, they  have lately made  repeated appearances 
in the context of integrable 
theories, e.g. in the  
$S$-matrices of affine Toda theories \BCDS, 
or in the excitation spectrum of integrable lattice models 
\MCOBSS.  


\Omit{
\vbox{
\subsec{The case of $c<1$ minimal models}
\nind
Closely related to WZW $sl(2)$ models.
If $c=1-{6(h-g)^2\over hg} $, $h,g\in {\IN}$,
irreps  $(r,s)\in \calI$ = ``Kac table'',
$1\le r\le h-1$, $1\le s\le g-1$,$(r,s)\equiv (h-r,g-s) $
and classification of modular invariants by
pairs of Dynkin diagrams $(A_{h-1},G)$, with $g$= Coxeter $\#$ of $G$.
\smallskip
\noindent
{\bf Problem:} Classify all ${\IN}$-valued matrix reps of minimal fusion
algebra.
\smallskip
\noindent
{\bf Theorem: } Only complete orthonormal solution to Cardy's equation
labelled by pairs $(r,a)$ of nodes of the $A_{h-1}$ and of
the $G$ graph, with the identification
$$
(r,a) \equiv (h-r,\gamma(a)) 
$$
where $\gamma$ is the automorphism of the $G$ Dynkin diagram,
(the natural  ${\IZ}_2$ symmetry for the $A$, $D_{{\rm odd}}$
and $E_6$ cases, the identity for the others)
$$
n_{rs}=N_r\otimes V_s + N_{h-r}\otimes V_{g-s}
$$
or
$$
n_{rs; (r_1,a)}{}^{(r_2,b)}=N_{rr_1}{}^{r_2}V_{sa}{}^b
+N_{h-r\,r_1}{}^{r_2}V_{g-s\,a}{}^b\ , 
$$
with $1\le r,r_1,r_2 \le h-1=2p$, $1\le s\le g-1$, and
$a,b$ running over the nodes of the Dynkin diagram $G$,
$N_r$: fusion matrices of $\widehat{sl}(2)_{h-2}$,
$V_s$: representation matrices of fusion algebra of
$\widehat{sl}(2)_{g-2}$ associated with graph $G$.
\smallskip
\noindent{\it Hints} Look at the generators $n_{21}$
and $n_{12}$ of fusion algebra. Both have eigenvalues
$|\gamma|<2$ \dots\dots
\smallskip
\noindent{\bf Example: the 3-state Potts model}\\
$c=4/5$, $(A_4,D_4)$
There are 8 distinct boundary consitions labelled by
$$(r,a)\in (A_4,D_4)/{\IZ}_2=(T_2,D_4)$$
cf Affleck-Ishikawa-Saleur
}} 

\subsec{Other cases}
\nind
It should be  clear that the situation that we have described 
in detail for $sl(2)$ extends to all RCFTs. The matrices $n_i$ 
solutions to Cardy equation are the adjacency matrices of graphs. In the 
case of $\slh(N)$, it is sufficient to supply the $(N-1)$ fundamental
matrices $\npbox$, 
$p=1,\cdots, N-1$,  to determine
all of them. The fact that  all $n_i$ then 
have non negative integer elements is
non trivial. By Cardy equation again, they satisfy a very restrictive
spectral property: their eigenvalues must be of the form $S_{ij}/S_{1j}$,
when $j$ runs over the set $\CE$, i.e. the diagonal part of the modular
invariant. We have thus found a justification of the 
empirical association between RCFTs and graphs, (see sect 2.4),
 and we have found
a physical interpretation of the matrix element $n_{ia}{}^b$ as the 
multiplicity of representation $i$ in the presence of the 
boundary conditions $a$ and $b$. 

The program of classifying these graphs/boundary conditions 
has been completed only in a few cases: $\slh(2)$ as discussed
above, $\slh(3)$ through a
combination of Gannon's work and the recent work of Ocneanu \AO, 
see sect. 2.3;
$\slh(N)_1$ \BPPZ, where the results match those obtained in 
the study of modular invariants \IDG: the graphs turn out to be star 
polygons.  The case of minimal ($c<1$) models has also been fully
analysed \BPPZ. 
%

\subsec{ Other algebraic features}

\nind
The identification of the allowed boundary conditions with the 
determination of the multiplicities and of the 
associated graphs is just the beginning of the story. A
 more elaborate discussion of BCFT should include a study of
the operator algebra in the presence of a boundary. This is an important, 
interesting and lively subject, a general understanding of which is
still missing. It is also beyond the scope of these introductory 
lectures. Let me only mention that in that study, 
 it appears that  a triplet of algebras $(n_i, M_i, N_a)$
plays a key role. Let  
\eqnn\troisalg
$$\eqalignno{ n_{ia}{}^b&=\sum_{\ell\in\calE} {S_{i\ell}\over S_{1\ell}}\;\psi_a^\ell
\left(\psi_b^\ell\right)\!{}^*
\cr 
M_{ij}{}^k &= \sum_{a\in G} {\psi_a^i \psi_a^j(\psi_a^k)^*\over
\psi_a^1} 
& \troisalg\cr
\hat N_{ab}{}^c &= \sum_{\ell\in \calE} {\psi_a^\ell \psi_b^\ell
(\psi_c^\ell)^*\over \psi_1^\ell}\ . 
\cr
}$$
\def\hN{\hat N}
The first has already been encountered. In the definition of the 
second (``Pasquier algebra''), the positivity of the components of the
Perron-Frobenius 
eigenvector $\psi^1$ is crucial. For the third, one assumes as above the 
existence of a special vertex denoted 1, such that all $\psi_1{}^\ell\ne 0$.
Note that as matrices, the $n$, $M$ and $\hN$ satisfy 
\eqnn\triplalg
$$\eqalignno{ n_i n_j&= \sum_k \CN_{ij}{}^k n_k\cr
M_i M_j &=\sum_k M_{ij}{}^k M_k & \triplalg\cr
\hN_a \hN_b &= \sum_c \hN_{ab}{}^c \hN_c \ .\cr
}$$
The role of the first has just been discussed.
In most known  cases the $\hN_{ab}{}^c$ turn out to be integers. 
In the type I cases, where they are non negative,  they
seem to describe a certain fusion algebra, 
generalised to a class of a yet ill-defined class of ``twisted''
 representations of the extended algebra. Restricted to 
the subset $a\in T$ (see sect. 3.6 (1)), it reduces to the ordinary fusion 
of ordinary representations of the extended algebra. 
One may show that the graph adjacency matrices $n_i$ are linear 
combinations of the $\hN_a$, and the algebra of the latter may be 
called a graph fusion algebra.
This graph fusion algebra 
is the dual of the Pasquier algebra in the sense
of the theory of {\it C-algebras} \BI,\DFZdeu,\PZdeu. 
In the context of BCFT, the Pasquier algebra is deeply connected
with the properties of the so-called 
bulk-boundary coefficients, which describe the 
coupling of bulk and boundary operators. 
For more details, I refer the interested 
reader to \BPPZ\ and to the many references quoted there.

I have been particularly sloppy on references in this last section. I should
mention that over the last ten years, this subject of boundary CFT has 
received important contributions by many, in particular Saleur and Bauer,
Cardy and Lewellen, Affleck and Ludwig, Affleck and 
Oshikawa, and Saleur, Pradisi, Sagnotti and Stanev, Recknagel and Schomerus,
Fuchs and Schweigert, and Runkel: these references may be found in \BPPZ.
Additional  recent references are by Huiszoon, Schellekens and  Sousa, 
by Gannon, by Felder, Fr\"ohlich, Fuchs and Schweigert, and many others.

\newsec{Lattice integrable realizations}

\nind It should also be mentionned that parallel to the conformal
field theoretic discussion sketched in these notes, there
exists a discussion of lattice integrable models, the so-called
face, or height, or RSOS, models. There the Yang-Baxter is realised 
through a representation of the Temperley-Lieb algebra, or of some
other quotient of the Hecke algebra, on the space of paths on a graph:
for example the Pasquier models \VPun\ in the simplest case of
$sl(2)$, or their higher rank generalizations. Finally, boundaries
may be introduced without spoiling integrability, through a careful
determination of the boundary Boltzmann weights, satisfying the
Boundary Yang-Baxter Equation \BP. 

Through these different approaches we can see the 
various facets of a beautiful common algebraic structure\dots

\bigskip\noindent{\bf Acknowledgements}\par\noindent
It is a pleasure to thank the two Roberts, Robert Coquereaux and 
Roberto Trinchero,  for their invitation and
hospitality in  beautiful Patagonia and for offering me the 
opportunity of extremely stimulating talks with many participants, 
in particular Adrian Ocneanu. I also want to thank Robert Coquereaux
and  David Evans for comments on the first draft,  and Valya Petkova 
for a careful and critical reading of these lecture notes and 
for thoughtful suggestions. 


\appendix{A}{The McKay correspondence} 

\def\bfz{0}
\def\bfa{{a}}\def\bfb{{b}}
\def\tta{{\bf a}}\def\ttb{{\bf b}}

The following presents details on the McKay correspondence, the 
Kostant polynomials and their connection with ``intertwiners''
of $ADE$ type.  The proof of the latter
 is a slightly expanded version of what appeared in a 
review article by Di Francesco \DFZdeu. 

Let $\Gamma$ be a subgroup of $SU(2)$. We denote by 
$(\bfa)$ its irreducible representations, among which ${(\bfz)}$
is the identity representation; $(f)$, the ``fundamental'',
is the two-dimensional representation of $\Gamma$ inherited from that
of $SU(2)$. (It may be irreducible or reducible, depending on
$\Gamma$). The fundamental observation of McKay \McK\ is that if we
tensor product $(\bfa)$ by $(f)$ and decompose it on irreps
\eqn\mck{(f)\otimes (\bfa) =\oplus_b \widehat{G}_{ab}\, (\bfb)}
we find that $\widehat{G}_{ab}$ is the adjacency matrix of an affine Dynkin
diagram $\widehat{G}$, thus canonically associated with $\Gamma$,
according to Table 3 of sect 2.2. In the following, we shall restrict to 
the subgroups $\Gamma$ such that $\hat G$ is bi-colourable, namely
$\CC_{2n},\CD_n,\CT,\CO,\CI$.

We want to see how the irreps of $SU(2)$ decompose onto the irreps
of $\GG$: let $[n]$ be the $(n+1)$-dimensional representation
of $SU(2)$ restricted to $\GG$ (in particular $[0]$ is the identity
representation, and $[1]=(f)$, see above. Let
\eqn\multip{[n]=\oplus _a N_{na}\, (\bfa)}
with a generating function of the multiplicities $N$ written as
\eqn\genfn{\eqalign{F(t)&=\sum_{n=0}^\infty \,t^n\, [n]
= \sum_{\bfa,n} t^n N_{na}\, (\bfa) \cr
&= \sum_{\bfa\atop {\rm irreps\ of\ }\GG} F_\bfa(t)\, (\bfa)\ .\cr}}
One writes easily recursion formulae
\eqn\recform{\eqalign{
[1]\otimes F(t)
&=\sum_\bfa F_\bfa(t) \, (\bfa) \otimes (f)\cr
&= \sum_{n=0}^\infty t^n\,\( [n+1]+[n-1]\) 
= \(t+{1\over t}\) F(t) -{(0)\over t}\cr
\ .}}
We evaluate this by taking its character on conjugation classes $C_i$
of $\GG$:
\eqn\eval{\sum_\bfa F_\bfa(t) \chi_\bfa(C_i) \chi_f(C_i) =
\(t+{1\over t}\)  \sum_\bfa F_\bfa(t) \chi_\bfa(C_i) -{1\over t}}
hence
\eqn\evalu{\sum_\bfa F_\bfa(t) \chi_\bfa(C_i)
={1\over 1+t^2 -t \chi_f(C_i)}\ ,}
or, using the orthogonality of characters $\sum_i\, |C_i|\,\chi_a(C_i) \chi_b^*(C_i)
=|\GG|\, \delta_{ab}$:
\eqn\Fsuba{F_\bfa(t)=
\sum_i {|C_i|\over |\GG|}\, {\chi_a^*(C_i)\over 1-t \chi_f(C_i)+t^2}
\ .}
The explicit result has been worked out by Kostant \Kos. He found
\eqn\Fakos{F_\bfa(t)={p_\bfa(t)\over (1-t^\tta)(1-t^\ttb)}}
where $p_\bfa(t)$ is a polynomial in $t$ of degree less or equal to $ h$ 
($h$ is the Coxeter number
of the {\it finite} Dynkin diagram $G$ associated with $\widehat{G}$,
$\tta$ and $\ttb$ are two integers satisfying $ \tta \ttb=2|\GG|$,
$\tta+\ttb=h+2$, for example 6 and 8 for $E_6$).
\par\nind{\Ex : prove that in \Fsuba\ only $|\Gamma|$-th roots of
unity appear as poles in $t$, 
hence that $\tta$ and $\ttb$ must divide $|\GG|$.}
\bigskip
\hrule
{ \halign{ # & # & # & # & # & # \cr
$\quad\qquad\Gamma$ & \qquad $\CC_{2n}$ &\qquad  $\CD_n$ &\qquad $\CT$& \qquad $\CO$ & \qquad $\bar\CI$\cr
$\quad\qquad|\Gamma|$ & \qquad $2n$ & \qquad  $4n$ & \qquad $24$ & \qquad $48$ &\qquad $120$ \cr
\quad\qquad$({\tta},{\ttb})$ & \qquad$(2,2n)$ & \qquad$(4,2n)$ &\qquad $(6,8)$ & \qquad$(8,12)$ & \qquad$(12,20)$ \cr
\qquad\quad$h$ &\qquad $2n$ & \qquad$2n+2$ & \qquad$12$ &\qquad $18$ & \qquad$30$ \cr
\quad\qquad$G$ &\ \ \ $\, A_{2n-1}$ &\  \ \  $\, D_{n+2}$ &\qquad  $E_6$ &\qquad $E_7$ &\qquad $E_8$ \cr
}
\smallskip
\hrule
\medskip

Let us plug the form above in the recursion formula, getting rid
of the denominator $(1-t^\tta)(1-t^\ttb)$
\eqnn\getrid
$$\eqalignno{\(t+{1\over t}\) \sum_\bfa p_\bfa(t) (\bfa) -{(0)\over t}
(1-t^\tta)(1-t^\ttb)&=\sum_\bfa p_\bfa(t) \, (\bfa) \otimes [1]\cr
&= \sum_{a,b} p_b(t) \widehat{G}_{ab} \, (a)\ . & \getrid\cr}$$
Denote by $G_{ab}$ the adjacency matrix of the {\it ordinary}
Dynkin diagram, obtained from $\widehat{G}$ by removing the node
called $\bfz$. Then
identifying in \getrid\ the coefficient of 
$(a)\ne(\bfz)$ 
gives
\eqn\coefbfa
{\bfa\ne \bfz\qquad \(t+{1\over t}\)  p_{\bfa}(t) 
= \sum_{\bfb\ne \bfz} p_\bfb(t) G_{ba}+p_0(t) \widehat{G}_{a0} 
\ . 
}
Following Kostant, write $p_\bfa(t)=\sum_{n=0}^{h} p_{\bfa\,n }\, t^n$,
$p_{a\,n}=p_{a\,h-n}$, $p_0(t)=1+t^h$. Eq. \coefbfa
\ implies that $ p_{\bfa\,0 }=p_{\bfa\,h }=0$ for $\bfa \ne \bfz$,
and thus
\eqn\recurs{\eqalign{
\(t+{1\over t}\)  p_{\bfa}(t) & =\sum_{n=1}^{h-1}  p_{\bfa\,n }\,
(t^{n+1}+t^{n-1}) \cr
&= \sum_{n=0}^{h}  (p_{\bfa\,n+1 } +p_{\bfa\,n-1 } )\, t^{n}
\qquad {\rm with}\ p_{\bfa\, -1}=p_{\bfa\, h+1}\equiv 0\cr
&= \sum_{m,n=1}^{h-1} t^n A_{nm} p_{\bfa\, m}+ t^0 p_{\bfa\, 1}
+t^h p_{\bfa\, h-1}\cr}}
where $A_{nm}$ is the adjacency matrix of type $A$. 
Compare the r.h.s. of equations \coefbfa\ and \recurs. 
Since degree$(p_\bfa)< h$ for $a\ne 0$, the two extra terms
in \recurs\ have to be identified with 
$p_\bfz(t)\widehat{G}_{a0}= (1+t^h)\widehat{G}_{a0}$ in \coefbfa, hence
\eqn\bcint{p_{\bfa\, 1}=p_{\bfa\, h-1}=\widehat{G}_{a0}\ ,}
while the identification of the coefficient of the terms 
$t^n$, $1\le n\le h-1$ yields
\eqn\entrelac{\sum_{m=1}^{h-1} A_{nm} p_{\bfa\, m}=
\sum_{\bfb \ne \bfz}  p_{\bfb\, n} G_{ba}\ .}
This may be read as a recursion formula determining uniquely   
$p_{a\,n+1}=p_{a\,n-1}+\sum_b G_{ba} p_{b\,n}$, starting from $p_{a\,1}=
\widehat{G}_{a0}$. 
But this also proves that $p_{\bfa\, m}$ is an intertwiner
between the $A$ and the $G$ Dynkin diagrams.
Its expression in terms of  the matrices of sect. 3.6
\eqn\identi{p_{\bfa\, m} = 
\sum_b \widehat{G}_{0b} n_{mb}{}^a }
follows from the observation that both satisfy the
boundary conditions \bcint.

This completes our proof of the statement (3) in sect 3.6, namely that
Kostant polynomials are given by the intertwiners $n$:
\eqn\Kostint{p_a(t)= \sum_{m=1}^{h-1}\sum_b \widehat{G}_{0b}
 n_{m\,b}{}^a\, t^m\ ,\qquad a\ne 0\ .}
%
\appendix{B}{The counting of essential paths} 

The following gives a slight variant of a proof given by
Ocneanu \Ocn, that the entry $n_{na}{}^b$ of the intertwiners
of $ADE$ type gives the number of ``essential paths'' on 
the diagram under consideration.

Given a graph of $ADE$ type, consider the set of paths
of length $n\ge 0$ starting from the vertex $a$ and ending at
vertex $b$,
$ \(a_0=a, a_1, \cdots, a_n=b\)$.
Consider the linear span $\CP_{ab}^{(n)} $  of these paths.
Define the operator $\GD_i$, the {\it contraction operator at step $i$}, by
\eqn\Bun
{\GD_i \(a_0=1, a_1, \cdots, a_n\)= 
\(a_0=1, a_1, \cdots, a_{i-1}, a_{i+2}, \cdots a_n\) \Gd_{a_{i-1},a_{i+1}}}
i.e. $\GD_i$ gives zero if the path doesn't backtrack.
Each $\GD_i$ maps $\CP^{(n)}_{ab}$ into $\CP^{(n-2)}_{ab}$.
Then define the   subspace $\CE_{ab}^{(n)}$ of essential paths
from $a$ to $b$ of length $n$ as those that are in the kernel
of all $\GD_i$ in $\CP_{ab}^{(n)}$, for $i=1,\cdots n-1$.
It is important that this concept is defined in
the vector space, because for two paths
that both  backtrack and would yield the same contracted path,
their {\it difference} is in the kernel. This is typically what
happens at the ``fork'' of a graph  $D$: the two paths that ``bounce''
off the two end points have a difference that is essential.
(As a side remark, it is more suitable to normalise differently the
contraction operator \Ocn. This does not affect the dimension of its 
kernel but the explicit form of its null vectors is changed.)

As proved by Ocneanu \Ocn,  the counting of essential paths of length $n$
with fixed ends $a,b$ (i.e. the dimension of $\CE_{ab}^{(n)}$ )
is given by the intertwiner $n_{n+1\,a}{}^b$.
In particular  the length of essential paths is bounded
by the Coxeter number $-1$. I give here a proof slightly different
from that of Ocneanu, in which I show that 
the recursion formula for essential paths is just the same
as that for the intertwiners, namely \recurr.

Denoting by $a:b$ the property of $a$ and $b$ to be 
neighbours on the graph, one has
\eqnn\Bdeu
$$\eqalignno{ \CE_{ab}^{(n)}&=
\cap_{i=1}^{n-1} (\ker \GD_i)_{\CP_{ab}^{(n)}} \cr
&= \Big(\oplus_{c:b} \cap_{i=1}^{n-2}(\ker \GD_i)_{\CP_{ac}^{(n-1)}}\Big) \cap
(\ker \GD_{n-1})_{\CP_{ab}^{(n)}} & \Bdeu \cr
&=(\ker \GD_{n-1})_{\oplus_{c:b}
\cap_{i=1}^{n-2}(\ker \GD_i)_{\CP_{ac}^{(n-1)}} } \ .\cr
}$$
Then consider  the {\it image of } $\GD_{n-1}$ in the space
$\oplus_{c:b} \cap_{i=1}^{n-2} (\ker \GD_i)_{\CP_{ac}^{(n-1)}} $
i.e. in the subspace of ${\CP_{ab}^{(n)}}$ of paths essential up to site
$n-2$ (the paths of those kernels are continued from
length $n-1$ to length $n$ by`` adding'' the last step $c-b$). This subspace
is $\CE^{(n-2)}_{ab}$, since  for a linear combination of paths $p$
\eqn\Btroi{ \GD_{n-1} \sum_p c_p [p=(a_0, \cdots, a_{n-1}(p), a_n=b)]
=\sum_p c_p  (a_0,\cdots, a_{n-2}(p)) \Gd_{a_{n-2},b}}
which is a generic element of $\CE_{ab}^{(n-2)}$. Hence
the dimension of its kernel in that space
i.e. the dimension of the r.h.s. of \Bdeu\ is
\eqn\Bquat{\CN_{ab}^{(n)}= \dim\CE^{(n)}_{ab}=\sum_{c:b}\dim\CE^{(n-1)}_{ac}
-\CN^{(n-2)}_{ab}\ .}
Thus the numbers $\CN$ satisfy the same recursion formula \recurr\
as the $n$'s, and the same boundary conditions, hence are identical. QED

\listrefs

\bye